\documentclass[english,reqno]{scrartcl}
\usepackage[T1]{fontenc}
\usepackage[latin9]{inputenc}
\usepackage{geometry}
\usepackage{color}
\usepackage{babel}
\usepackage{units}
\usepackage{textcomp}
\usepackage{amsmath}
\usepackage{amssymb}
\usepackage{esint}
\usepackage[unicode=true,pdfusetitle,
 bookmarks=true,bookmarksnumbered=false,bookmarksopen=false,
 breaklinks=false,pdfborder={0 0 0},backref=false,colorlinks=true]
 {hyperref}
\usepackage{breakurl}

\makeatletter
\numberwithin{equation}{section}
\numberwithin{figure}{section}

\usepackage{graphicx}
\usepackage{fancybox}
\usepackage{url}
\usepackage{hyperref}

\makeatother

\begin{document}
\global\long\def\tl#1{\dot{#1}}
\global\long\def\part#1#2{\frac{\partial#1}{\partial#2}}
\global\long\def\adj#1{\bar{#1}}
\global\long\def\p#1{\partial#1}
\global\long\def\Va{V_{\alpha}}
\global\long\def\Vb{V_{\beta}}
\global\long\def\Vd{V_{\delta}}
\global\long\def\Vbmax{\bar{V}_{MAX}}
\global\long\def\l#1{\lambda_{#1}}
\global\long\def\d{\delta}
\global\long\def\a{\alpha}
\global\long\def\b{\beta}
\global\long\def\g{\gamma}
\global\long\def\amin{\alpha_{MIN}}
\global\long\def\bmin{\beta_{MIN}}
\global\long\def\gmin{\gamma_{MIN}}
\global\long\def\Vbsq#1{\bar{V}_{#1}}
\global\long\def\mktvolsq#1{\left[\sigma_{#1}^{MKT}\right]^{2}}
\global\long\def\bsm{\mathcal{BSM}}

\title{Implied volatility surface: construction methodologies and characteristics }

\author{Cristian Homescu%
\thanks{Email address: cristian.homescu@gmail.com%
}}

\date{This version: July 9, 2011%
\thanks{Original version: July 9, 2011%
}}

\maketitle
The implied volatility surface (IVS) is a fundamental building block
in computational finance. We provide a survey of methodologies for
constructing such surfaces. We also discuss various topics which influence
the successful construction of IVS in practice: arbitrage-free conditions
in both strike and time, how to perform extrapolation outside the
core region, choice of calibrating functional and selection of numerical
optimization algorithms, volatility surface dynamics and asymptotics.

\pagebreak{}

\tableofcontents{}

\pagebreak{}

\section{Introduction}

The geometric Brownian motion dynamics used by Black and Scholes (1973)
and Merton (1973) to price options constitutes a landmark in the development
of modern quantitative finance. Although it is widely acknowledged
that the assumptions underlying the Black-Scholes-Merton model (denoted
$\bsm$ for the rest of the paper) are far from realistic, the $\mbox{\ensuremath{\bsm}}$
formula remains popular with practitioners, for whom it serves as
a convenient mapping device from the space of option prices to a single
real number called the implied volatility (IV). This mapping of prices
to implied volatilities allows for easier comparison of options prices
across various strikes, expiries and underlying assets.

When the implied volatilities are plotted against the strike price
at a fixed maturity, one often observes a skew or smile pattern, which
has been shown to be directly related to the conditional non-normality
of the underlying return risk-neutral distribution. In particular,
a smile reflects fat tails in the return distribution whereas a skew
indicates return distribution asymmetry. Furthermore, how the implied
volatility smile varies across option maturity and calendar time reveals
how the conditional return distribution non-normality varies across
different conditioning horizons and over different time periods. For
a fixed relative strike across several expiries one speaks of the
term structure of the implied volatility.

We mention a few complications which arise in the presence of smile.
Arbitrage may exist among the quoted options. Even if the original
market data set does not have arbitrage, the constructed volatility
surface may not be arbitrage free. The trading desks need to price
European options for strikes and maturities not quoted in the market,
as well as pricing and hedging more exotic options by taking the smile
into account. 

Therefore there are several practical reasons \cite{Fengler_2010}
to have a smooth and well-behaved implied volatility surface (IVS):
\begin{enumerate}
\item market makers quote options for strike-expiry pairs which are illiquid
or not listed; 
\item pricing engines, which are used to price exotic options and which
are based on far more realistic assumptions than $\bsm$ model, are
calibrated against an observed IVS; 
\item the IVS given by a listed market serves as the market of primary hedging
instruments against volatility and gamma risk (second-order sensitivity
with respect to the spot); 
\item risk managers use stress scenarios defined on the IVS to visualize
and quantify the risk inherent to option portfolios.
\end{enumerate}
The IVS is constructed using a discrete set of market data (implied
volatilities or prices) for different strikes and maturities. Typical
approaches used by financial institutions are based on:
\begin{itemize}
\item (local) stochastic volatility models
\item Levy processes (including jump diffusion models)
\item direct modeling of dynamics of the implied volatility
\item parametric or semi-parametric representations
\item specialized interpolation methodologies 
\end{itemize}
Arbitrage conditions may be implicitly or explicitly embedded in the
procedure

This paper gives an overview of such approaches, describes characteristics
of volatility surfaces and provides practical details for construction
of IVS.

\section{Volatility surfaces based on (local) stochastic volatility models}

A widely used methodology employs formulae based from stochastic volatility
models to fit the set of given market data. The result is an arbitrage
free procedure to interpolate the implied volatility surface. The
most commonly considered stochastic volatility models are Heston and
SABR and their extensions (such as time dependent parameters, etc)
and we will concentrate on these models as well. Having time dependent
parameters allows us to perform calibration in both strike and time
directions. This is arguably better than the case of using constant
parameter models in capturing inter-dependencies of different time
periods. The main disadvantage when using time dependent parameters
is the increase in computational time, since in many cases we do not
have analytical solutions/approximations and we have to resort to
numerical solutions when performing the calibration. However, for
the considered time dependent models, namely Heston and SABR, (semi)analytical
approximations are available, which mitigates this issue. 

We will also consider the hybrid local stochastic volatility models,
which are increasingly being preferred by practitioners, and describe
efficient calibration procedures for such models.

\subsection{Heston model and its extensions }

The Heston model is a lognormal model where the square of volatility
follows a Cox\textendash{}Ingersoll\textendash{}Ross (CIR) process.
The call (and put) price has a closed formula through to a Fourier
inversion of the characteristic function. Details on efficient ways
to compute those formulas are given in \cite{Kilin_2011}, recent
advances were presented in \cite{Jacquier_2010}, while \cite{Gauthier_Rivaille_2009}
contains details for improving the numerical calibration.

It is our experience, confirmed by discussions with other practitioners,
that having only one set of parameters is usually not enough to match
well market data corresponding to the whole range of expiries used
for calibration, especially for FX markets. Consequently we need to
consider time dependent parameters.

When the parameters are piecewise constant, one can still derive a
recursive closed formula using a PDE/ODE methodology \cite{Mikhailov_Nogel_2004}
or a Markov argument in combination with affine models \cite{Elices_2008},
but formula evaluation becomes increasingly time consuming. 

A better and more general approach is presented in \cite{Benhamou_et_al_2010},
which is based on expanding the price with respect to the volatility
of volatility (which is quite small in practice) and then computing
the correction terms using Malliavin calculus. The resulting formula
is a sum of two terms: the $\bsm$ price for the model without volatility
of volatility, and a correction term that is a combination of Greeks
of the leading term with explicit weights depending only on the model
parameters. 

The model is defined as 
\begin{eqnarray}
dX(t) & = & \sqrt{\nu(t)}dW_{1}(t)-\frac{\nu(t)}{2}dt\label{eq:Heston}\\
d\nu(t) & = & \kappa(t)\left(\theta(t)-\nu(t)\right)dt+\xi(t)\sqrt{\nu(t)}dW_{2}(t)\nonumber \\
d\left\langle W_{1},W_{2}\right\rangle  & = & \rho(t)dt\nonumber 
\end{eqnarray}

with initial conditions
\begin{eqnarray}
X(0) & = & x_{0}\nonumber \\
\nu(0) & = & \nu_{0}\label{eq:HestonIC}
\end{eqnarray}

Using the same notations as in \cite{Benhamou_et_al_2010}, the price
for the put option is 
\[
Put(K,T)=\exp\left[-\int_{0}^{T}r(t)dt\right]\mathbb{E}\left[\left(K-\exp\left[-\int_{0}^{T}\left(r(t)-q(t)\right)dt+X(T)\right]^{+}\right)\right]
\]

where $r(t),q(t)$ are the risk free rate and, respectively, dividend
yield, K is the strike and T is the maturity of the option.

There are two assumptions employed in the paper: 

1) Parameters of the CIR process verify the following conditions 
\begin{eqnarray*}
\xi_{inf} & > & 0\\
\left(\frac{2\kappa\theta}{\xi^{2}}\right)_{inf} & \geq & 1
\end{eqnarray*}

2) Correlation is bounded away from -1 and 1 
\[
\left\Vert \rho_{sup}\right\Vert <1
\]

Under these assumptions, the formula for approximation is 
\[
P_{BS}\left(x_{0},var(T)\right)+\sum_{i=1}^{2}a_{i}(T)\frac{\partial^{(i+1)}P_{BS}}{\partial x^{i}\partial y}\left(x_{0},var(T)\right)+\sum_{i=1}^{2}b_{2i}(T)\frac{\partial^{(2i+2)}P_{BS}}{\partial x^{2i}\partial y^{2}}\left(x_{0},var(T)\right)
\]

where $P_{BS}(x,y)$ is the price in a $\bsm$ model with spot $e^{x}$,
strike K, total variance $y$, maturity T and rates $r_{eq},q_{eq}$
given by 
\begin{eqnarray*}
r_{eq} & = & \frac{\int_{0}^{T}r(t)dt}{T}\\
q_{eq} & = & \frac{\int_{0}^{T}q(t)dt}{T}
\end{eqnarray*}

while $var(T),a_{i}(T),b_{2i}(T)$ have the following formulas 
\begin{eqnarray*}
var(T) & = & \intop_{0}^{T}\nu_{0}(t)dt\\
a_{1}(t) & = & \intop_{0}^{T}e^{\kappa s}\rho(s)\xi(s)\nu_{0}(s)\left(\intop_{s}^{T}e^{-\kappa u}du\right)ds\\
a_{2}(t) & = & \intop_{0}^{T}e^{\kappa s}\rho(s)\xi(s)\nu_{0}(s)\left(\intop_{s}^{T}\rho(t)\xi(t)\left(\intop_{t}^{T}e^{-\kappa u}du\right)dt\right)ds\\
b_{0}(t) & = & \intop_{0}^{T}e^{2\kappa s}\xi^{2}(s)\nu_{0}(s)\left(\intop_{s}^{T}e^{-\kappa t}\left(\intop_{t}^{T}e^{-\kappa u}du\right)dt\right)ds\\
b_{2}(T) & = & \frac{a_{1}^{2}(T)}{2}\\
\nu_{0}(t) & \triangleq & e^{-\kappa t}\left(\nu_{0}+\intop_{0}^{T}\kappa e^{\kappa s}\theta(s)ds\right)\\
\end{eqnarray*}

The error is shown to be of order $\mathcal{O}\left(\xi_{sup}^{3}T^{2}\right)$

While results are presented for general case, \cite{Benhamou_et_al_2010}
also includes explicit formulas for the case of piecewise constant
parameters. Numerical results show that calibration is quite effective
and that the approximation matches well the analytical solution, which
requires numerical integration. They report that the use of the approximation
formula enables a speed up of the valuation (and thus the calibration)
by a factor 100 to 600. 

We should also mention the efficient numerical approach presented
in \cite{Maruhn_2011} for calibration of the time dependent Heston
model. The constrained optimization problem is solved with an optimization
procedure that combines Gauss-Newton approximation of the Hessian
with a feasible point trust region SQP (sequential quadratic programming)
technique developed in \cite{Wright_Tenny_2004}. As discussed in
a later chapter on numerical remarks for calibration, in the case
of local minimizer applied to problems with multiple local/global
minima, a regularization term has to be added in order to ensure robustness/stability
of the solution.

\subsection{SABR model and its extensions }

The SABR model \cite{Hagan_et_al_2002} assumes that the forward asset
price $F(t)$ and its instantaneous volatility $\alpha(t)$ are driven
by the following system of SDEs: 
\begin{eqnarray}
dF(t) & = & \alpha(t)F^{\beta}(t)dW_{1}(t)\label{eq:sabr}\\
d\alpha(t) & = & \nu\alpha(t)dW_{2}(t)\nonumber \\
d\left\langle W_{1},W_{2}\right\rangle  & = & \rho dt\nonumber 
\end{eqnarray}

where is $\nu>0$ is volatility of volatility and $\beta>0$ is a
leverage coefficient. The initial conditions are 
\begin{eqnarray*}
F(0) & = & F_{0}\\
\alpha(0) & = & \alpha_{0}
\end{eqnarray*}

Financial interpretation for this model is the following: $\alpha(t)$
determines the overall level of at-the-money forward volatility; $\beta$
measures skew with two particular choices: $\beta=1$ corresponding
to the log-normal model with a flat skew and $\beta=0$ corresponding
to the normal model with a very pronounced skew; $\rho$ also controls
the skew shape with the choice $\rho<0$ (respectively $\rho>0$)
yielding the negative (respectively inverse) skew and with the choice
$\rho=0$ producing a symmetric volatility smile given $ $$\beta=1$;
$ $$\nu$ is a measure of convexity, i.e. stochasticity of $\alpha(t)$.

Essentially, this model assumes CEV distribution (log-normal in case
$\beta=1$) for forward price $F(t)$ and log-normal distribution
for instantaneous volatility $\alpha(t)$. 

SABR model is widely used by practitioners, due to the fact that it
has available analytical approximations. Several approaches were used
in the literature for obtaining such approximations: the singular
perturbation, heat kernel asymptotics, and Malliavin calculus \cite{Johnson_Nonas_2009,Obloj_2008,Hagan_et_al_2002}.
Additional higher order approximations are discussed in \cite{Paulot_2009}(second
order) and \cite{Takahashi_et_al_2009}, up to fifth order. Details
for improving the numerical calibration were given in \cite{Gauthier_Rivaille_2009}

An extension of SABR (termed lambda-SABR), and corresponding asymptotic
approximations were introduced in papers by Henry-Labordere (see chapter
6 of \cite{HenryLabordere_2008}). This model is described as follows
(and degenerates into SABR model when $\lambda=0$) 
\begin{eqnarray}
dF(t) & = & \alpha(t)F^{\beta}(t)dW_{1}(t)\label{eq:lambdaSABR}\\
d\alpha(t) & = & \lambda\left(\alpha(t)-\bar{\lambda}\right)+\nu\alpha(t)dW_{2}(t)\nonumber \\
d\left\langle W_{1},W_{2}\right\rangle  & = & \rho dt\nonumber 
\end{eqnarray}

The high order approximations in \cite{Takahashi_et_al_2009} were
obtained for lambda-SABR model. Approximations for extended lambda-SABR
model, where a drift term is added to the SDE describing $F(t)$ in
\eqref{eq:lambdaSABR}, were presented in \cite{Shiraya_Takahashi_2010}
and \cite{Shiraya_et_al_2009}. Approximations for SABR with time
dependent coefficients were presented in \cite{Osajima_2007}, where
the model was named \textquotedblleft{}Dynamic SABR\textquotedblright{},
and in respectively in \cite{Glasserman_Wu_2010}, where the approach
was specialized to piecewise constant parameters. 

If one combines the results from \cite{Shiraya_et_al_2009,Takahashi_et_al_2009,Shiraya_Takahashi_2010}
with the findings of \cite{Glasserman_Wu_2010}, the result will be
a model (extended lambda-SABR with piecewise constant parameters)
that may be rich to capture all desired properties when constructing
a volatility surface and yet tractable enough due to analytical approximations.

Alternatively, the results presented in \cite{Glasserman_Wu_2010}
seem very promising and will be briefly described below. The procedure
is based on asymptotic expansion of the bivariate transition density
\cite{Wu_2010}. 

To simplify the notations, the set of SABR parameters is denoted by
$\theta\triangleq\left\{ \alpha,\beta,\rho,\nu\right\} $ and the
dependence of the model\textquoteright{}s joint transition density
on the model parameters by $p\left(t,F_{0},\alpha_{0};T,\hat{F,}\hat{A};\theta\right)$.

The joint transition density is defined as 
\[
\mathbb{P}\left(\hat{F}<F(T)\leq\hat{F}+d\hat{F},\hat{A}\leq\alpha(T)\leq\hat{A}+d\hat{A}\right)\triangleq p\left(t,\hat{\bar{F}},\hat{\bar{A}};T,\hat{F},\hat{A}\right)d\hat{F}d\hat{A}
\]

We follow the notations from \cite{Wu_2010}, namely: 
\begin{itemize}
\item $\hat{F},\hat{A}$ are forward variables denoting the state values
of $F(T),\alpha(T)$
\item $ $$\hat{\bar{F}},\hat{\bar{A}}$ are backward variables denoting
the state values of $F(t),\alpha(t)$ 
\end{itemize}
Let us denote by $\left\{ T_{1},T_{2},...,T_{N}\right\} $ the set
of expiries for which we have market data we want to calibrate to;
we assume that the four SABR parameters $\left\{ \alpha,\beta,\rho.\nu\right\} $
are piecewise constant on each interval $\left[T_{i-1},T_{i}\right]$.

The tenor-dependent SABR model then reads 
\begin{eqnarray}
dF(t,T_{i}) & = & \alpha(t,T_{i})F^{\beta_{i}}(t)dW_{1}(t)\label{eq:TenorDependentSABR}\\
d\alpha(t,T_{i}) & = & \nu_{i}\alpha(t,T_{i})dW_{2}(t)\nonumber \\
\mathbb{E}^{\mathbb{Q}^{T_{i}}}\left[dW_{1}(t)dW_{2}(t)\right] & = & \rho_{i}dt\nonumber 
\end{eqnarray}

where $\mathbb{E}^{\mathbb{Q}^{T_{i}}}$ is the expectation under
the $T_{i}$ forward measure $\mathbb{Q}^{T_{i}}$ 

The SDE \eqref{eq:TenorDependentSABR} is considered together with
\begin{eqnarray*}
F(0,T_{i}) & = & F_{i}\\
\alpha(0,T_{i}) & = & \alpha_{i}
\end{eqnarray*}

The notations for SABR set of parameters and, respectively, for the
dependence of the model\textquoteright{}s joint transition density
on the model parameters are updated as follows
\[
\theta(T)=\begin{cases}
\left(\alpha_{0},\beta_{0},\rho_{0},\nu_{0}\right) & if\, T\leq T_{1}\\
\left(\alpha_{i-1},\beta_{i-1},\rho_{i-1},\nu_{i-1}\right) & if\, T_{i-1}<T\leq T_{i}\\
\left(\alpha_{N-1},\beta_{N-1},\rho_{N-1},\nu_{N-1}\right) & if\, T_{N-1}<T\leq T_{N}
\end{cases}
\]

and, respectively
\begin{eqnarray*}
p\left(0,F_{0},\alpha_{0};T_{1},F_{1};\theta_{0}\right)\\
p\left(T_{i-1},F_{i-1},A_{i-1};T_{i},F_{i},A_{i};\theta_{i-1}\right)\\
p\left(T_{N-1},F_{N-1},A_{N-1};T_{N},F_{N},A_{N};\theta_{N}\right)
\end{eqnarray*}

In the case where only the parameter dependence need to be stressed,
we use the shortened notion: 
\begin{eqnarray*}
p\left(0,T_{1};\theta_{0}\right)\\
p\left(T_{i-1},T_{i};\theta_{i-1}\right)
\end{eqnarray*}

A standard SABR model describes the dynamics of a forward price process
$F\left(t;T_{i}\right)$ maturing at a particular $T_{i}$. Forward
prices associated with different maturities are martingales with respect
to different forward measures defined by different zero-coupon bonds
$B\left(t,T_{i}\right)$ as numeraires. This raises consistency issues,
on both the underlying and the pricing measure, when we work with
multiple option maturities simultaneously. 

We address this issue by consolidating all dynamics into those of
$F\left(t,T_{N}\right),\alpha\left(t,T_{N}\right)$, whose tenor is
the longest among all, and express all option prices at different
tenors in one terminal measure $\mathbb{Q}^{T_{N}}$ which is the
one associated with the zero-coupon bond $B\left(t,T_{N}\right)$
. 

We may do so because we assume 

\textbullet{} No-arbitrage between spot price $S(t)$ and all of its
forward prices $F\left(t,T_{i}\right),i=1...N$, at all trading time
t; 

\textbullet{} Zero-coupon bonds $B\left(t,T_{i}\right)$~are risk-less
assets with positive values

Based on these assumptions we obtain the following formulas 
\begin{eqnarray*}
F\left(t,T_{1}\right) & = & \frac{S}{B\left(t,T_{1}\right)}\\
F\left(t,T_{i}\right) & =F\left(t,T_{N}\right) & \frac{B\left(t,T_{N}\right)}{B\left(t,T_{i}\right)}
\end{eqnarray*}

This will enable us to convert an option on $F\left(\cdot,T_{i}\right)$
into an option on $F\left(\cdot,T_{N}\right)$. The price of a call
option on $F\left(\cdot,T_{i}\right)$ with strike price $K_{j}$
and maturity $T_{i}$ then becomes 

\[
V\left(t,T_{i},K_{j}\right)=B\left(t,T_{N}\right)\mathbb{E}^{\mathbb{Q}^{T_{N}}}\left[\left(F\left(T_{i},T_{N}\right)-\bar{K}_{j}\right)^{+}|\Im_{t}\right]
\]

where the adjusted strike $\bar{K}_{j}$ is defined as 
\[
\bar{K}_{j}\triangleq\frac{K_{j}}{B\left(T_{i},T_{N}\right)}
\]

In the context of model calibration, computation of spot implied volatilities
from the model relies on computation of option prices 
\begin{equation}
\mathbb{E}^{\mathbb{Q}^{T_{N}}}\left[\left(F\left(T_{i}\right)-\bar{K}_{j}\right)^{+}|F_{i-1},A_{i-1}\right]=\iint_{\mathbb{R}_{+}^{2}}\left[\left(F\left(T_{i}\right)-\bar{K}_{j}\right)^{+}p\left(T_{i-1},T_{i};\theta_{i-1}\right)\right]dF_{i}dA_{i}\label{eq:OpionPriceSABRWu}
\end{equation}

at each tenor $T_{i}\,$ for each equivalent strike $\bar{K}_{j}$
. 

Asymptotic expansions of a more generic joint transition density have
been obtained analytically in \cite{Wu_2010} to the $n^{th}$ order.
We should note that for simplicity we use exactly the same notations
as in section 4.2 of \cite{Glasserman_Wu_2010}, namely that the values
of state variables at time $t$ are denoted by $f,\alpha$ and, respectively,
the values of the state variables at $T$ are denoted by $F,A$. 

The expansion to second order was shown to give a quite accurate approximation
\[
p_{2}\left(t,f,\alpha;T,F,A;\theta\right)=\frac{1}{\nu TF^{\beta}A^{2}}\left[\hat{p_{0}}+\nu^{2}\hat{p}_{1}\sqrt{T}+\nu^{2}\hat{p}_{2}T\right]\left(\tau,u,v,\theta\right)
\]

where we define $\tau,u,v$ as 
\begin{eqnarray*}
\tau & \triangleq & \frac{T-t}{T}\\
u & \triangleq & \frac{f^{1-\beta}-F^{1-\beta}}{\alpha\left(1-\beta\right)\sqrt{T}}\\
v & \triangleq & \frac{\ln\left(\frac{\alpha}{A}\right)}{\nu\sqrt{T}}
\end{eqnarray*}

The terms $\hat{p_{0}},\hat{p_{1}},\hat{p_{2}}$ have the following
formulae 
\begin{eqnarray*}
\hat{p_{0}}\left(\tau,u,v,\theta\right) & = & \frac{1}{2\pi\sqrt{1-\rho^{2}}}\exp\left[-\frac{u^{2}-2\rho uv+v^{2}}{2\tau\left(1-\rho^{2}\right)}\right]\\
\hat{p_{1}}\left(\tau,u,v,\theta\right) & = & \frac{a_{11}+\frac{a_{10}}{\tau}}{2\left(\rho^{2}-1\right)}\hat{p_{0}}\left(\tau,u,v,\theta\right)\\
\hat{p_{2}}\left(\tau,u,v,\theta\right) & = & \frac{a_{23}\tau+a_{22}+\frac{a_{21}}{\tau}+\frac{a_{20}}{\tau^{2}}}{24\left(1-\rho^{2}\right)^{2}}\hat{p_{0}}\left(\tau,u,v,\theta\right)
\end{eqnarray*}

Explicit expressions for the polynomial functions $a_{11},a_{10},a_{23},a_{22},a_{21},a_{20}$
are given in Eq. (42) in \cite{Wu_2010}. In terms of computational
cost, it is reported in \cite{Glasserman_Wu_2010} that it takes about
1-10 milliseconds for an evaluation of the integral \eqref{eq:OpionPriceSABRWu}
on a 1000 by 1000 grid, using the approximation $\hat{p_{2}}$ as
density.

\subsection{Local stochastic volatility model}

More and more practitioners are combining the strengths of local and
stochastic volatility models, with the resulting hybrid termed local
stochastic volatility (LSV) model. 

Based on \cite{Maruhn_2011}, we describe efficient procedures for
calibrating one such model, namely a hybrid Heston plus local volatility
model, with dynamics given by
\begin{eqnarray*}
df^{LSV}(t) & = & \sigma\left(f^{LSV}(t),t\right)\sqrt{v(t)}f^{LSV}(t)dW_{1}(t)\\
dv(t) & = & \kappa\left(\theta-v(t)\right)dt+\xi\sqrt{v(t)}dW_{2}(t)
\end{eqnarray*}

The calibration procedure is based on the following 2 step process:
\begin{itemize}
\item calibrate stochastic volatility component
\item perform LSV correction
\end{itemize}
The validity of this 2-step process is due to the observation that
the forward skew dynamics in stochastic volatility setting are mainly
preserved under the LSV correction.

The first approach is based on the {}``fixed point'' concept described
in \cite{Ren_et_al_2007}
\begin{enumerate}
\item Solve forward Kolmogorov PDE (in $x=\ln\left(\nicefrac{S}{fwd}\right)$
with a given estimate of $\sigma(f,t)$ 
\[
\part pt=\part{}x\left[\frac{1}{2}v\sigma^{2}p\right]-\part{}v\left[\kappa\left(\theta-v\right)p\right]+\part{^{2}}{x^{2}}\left[\frac{1}{2}v\sigma^{2}p\right]+\part{^{2}}{x\partial v}\left[\rho\sigma\xi vp\right]+\part{^{2}}{v^{2}}\left[\frac{1}{2}v\xi^{2}p\right]
\]

\item Use the density from 1. to compute the conditional expected value
of $v(t)$ given $f^{LSV}(t)$
\[
E\left[v(t)|f^{LSV}(t)=f\right]=\frac{\intop_{0}^{\infty}vp(t,f,v)dv}{\intop_{0}^{\infty}p(t,f,v)dv}
\]

\item Adjust $\sigma$ according to Gyongy's identity \cite{Gyongy_1986}
for the local volatilities of the LSV model
\[
\left(\sigma_{LV}^{LSV}\right)^{2}(f,t)=\sigma^{2}(f,t)E\left[v(t)|f^{LSV}(t)=f\right]=\left(\sigma_{LV}^{Market}\right)^{2}(f,t)
\]

\item repeat steps 1.-3. until $\sigma(f,t)$ has converged (it was reported
that in most cases 1-2 loops are sufficient)
\end{enumerate}
The second approach is based on {}``local volatility'' ratios, similar
to \cite{Piterbarg_2007,HenryLabordere_2009}. The main idea is the
following: applying Gyongy's theorem \cite{Gyongy_1986} twice (for
the starting stochastic volatility component and, respectively, for
the target LSV model) avoids the need for conditional expectations. 

The procedure is as follows
\begin{enumerate}
\item Compute the local volatilities of an LSV and an SV model via Gyongy's
formula
\begin{eqnarray*}
\left(\sigma_{LV}^{LSV}\right)^{2}(f,t) & = & \sigma^{2}(f,t)E\left[v(t)|f^{LSV}(t)=f\right]=\left(\sigma_{LV}^{Market}\right)^{2}(f,t)\\
\left(\sigma_{LV}^{SV}\right)^{2}(x,t) & = & E\left[v(t)|f^{SV}(t)=x\right]
\end{eqnarray*}

\item Taking the ratio and solving for the unknown function $\sigma(\cdot,\cdot)$
we obtain
\[
\sigma(t,f)=\frac{\sigma_{LV}^{Market}(f,t)}{\sigma_{LV}^{SV}(x,t)}\sqrt{\frac{E\left[v(t)|f^{SV}(t)=x\right]}{E\left[v(t)|f^{LSV}(t)=f\right]}}\thickapprox using\,\, x=H(f,t)\thickapprox\frac{\sigma_{LV}^{Market}(f,t)}{\sigma_{LV}^{SV}(H(f,t),t)}
\]
 with an approximate, strictly monotonically increasing map $H(f,t)$
\end{enumerate}
The calculation is reported to be extremely fast if the starting local
volatilities are easy to compute. The resulting calibration leads
to near perfect fit of the market

We should also mention a different calibration procedure for a hybrid
Heston plus local volatility model, presented in \cite{Engelmann_et_al_2011}.

\section{Volatility surfaces based on Levy processes}

Volatility surface representations based on Levy processes tend to
better handle steep short term skews (observed especially in FX and
commodity markets). In a model with continuous paths like a diff{}usion
model, the price process behaves locally like a Brownian motion and
the probability that the price of the underlying moves by a large
amount over a short period of time is very small, unless one fi{}xes
an unrealistically high value of volatility. Thus in such models the
prices of short term OTM options are much lower than what one observes
in real markets. On the other hand, if price of underlying is allowed
to jump, even when the time to maturity is very short, there is a
non-negligible probability that after a sudden change in the price
of the underlying the option will move in the money. 

The Levy processes can be broadly divided into 2 main categories:
\begin{itemize}
\item jump diffusion processes: jumps are considered rare events, and in
any given fi{}nite interval there are only fi{}nite many jumps
\item infinite activity Levy processes: in any fi{}nite time interval there
are infi{}nitely many jumps.
\end{itemize}
The importance of a jump component when pricing options close to maturity
is also pointed out in the literature, e.g., \cite{Andersen_Andreasen_2000}.
Using implied volatility surface asymptotics, the results from \cite{Medvedev_Scaillet_2006}
confi{}rm the presence of a jump component when looking at S\&P option
data .

Before choosing a particular parametrization, one must determine the
qualitative features of the model. In the context of Levy-based models,
the important questions are \cite{Cont_Tankov_2004,Tankov_2007}:
\begin{itemize}
\item Is the model a pure-jump process, a pure diff{}usion process or a
combination of both? 
\item Is the jump part a compound Poisson process or, respectively, an infi{}nite
intensity Levy process? 
\item Is the data adequately described by a time-homogeneous Levy process
or is a more general model may be required?
\end{itemize}
Well known models based on Levy processes include Variance Gamma \cite{Madan_et_al_1998},
Kou \cite{Kou_2002}, Normal Inverse Gaussian \cite{BarndorffNielsen_1998},
Meixner \cite{Schoutens_2002,Madan_Yor_2008}, CGMY \cite{Carr_et_al_2002},
affine jump diffusions \cite{Duffie_et_al_2000}.

From a practical point of view, calibration of Levy-based models is
definitely more challenging, especially since it was shown in \cite{Cont_Tankov_2004a,Cont_Tankov_2006}
that it is not sufficient to consider only time-homogeneous Levy specifi{}cations.
Using a non-parametric calibration procedure, these papers have shown
that Levy processes reproduce the implied volatility smile for a single
maturity quite well, but when it comes to calibrating several maturities
at the same time, the calibration by Levy processes becomes much less
precise. Thus successful calibration procedures would have to be based
on models with more complex characteristics.

To calibrate a jump-diff{}usion model to options of several maturities
at the same time, the model must have a suffi{}cient number of degrees
of freedom to reproduce diff{}erent term structures. This was demonstrated
in \cite{Tankov_Voltchkova_2009} using the Bates model, for which
the smile for short maturities is explained by the presence of jumps
whereas the smile for longer maturities and the term structure of
implied volatility is taken into account using the stochastic volatility
process.

In \cite{Galluccio_LeCam_2007} a stochastic volatility jump diffusion
model is calibrated to the whole market implied volatility surface
at any given time, relying on the asymptotic behavior of the moments
of the underlying distribution. A forward PIDE (Partial Integro-Differential
Equation) for the evolution of call option prices as functions of
strike and maturity was used in \cite{Andersen_Andreasen_2000} in
an effi{}cient calibration to market quoted option prices, in the
context of adding Poisson jumps to a local volatility model.

\subsection{Implied Levy volatility}

An interesting concept was introduced in \cite{Corcuera_at_al_2009},
which introduced the implied Levy space volatility and the implied
Levy time volatility, and showed that both implied Levy volatilities
would allow an exact fit of the market. Instead of normal distribution,
as is the case for implied volatility calculation, their starting
point is a distribution that was more in line with the empirical observations.

More specifically, instead of lognormal model they assume the following
model
\begin{equation}
S(t)=S_{0}\exp\left[\left(r-q+\omega\right)t+\sigma X(t)\right]\label{eq:sdeLevyImpliedVol}
\end{equation}

where $\sigma>0$, $r$ is the risk-free rate, $q$ is the dividend
yield, $\omega$ is a term that is added in order to make dynamics
risk neutral, and $X=\left\{ X(t),t\geq0\right\} $ is a stochastic
process that starts at zero, has stationary and independent increments
distributed according to the newly selected distribution

Once one has fixed the distribution of $X$ (which we assume as in
the Brownian case to have mean zero and variance at unit time equal
to 1), for a given market price one can look for the corresponding
$\sigma$, which is termed the implied (space) Levy volatility, such
that the model price matches exactly the market price.

To define Implied Levy Space Volatility, we start with an infinitely
divisible distribution with zero mean and variance one and denote
the corresponding Levy process by $X=\left\{ X(t),t\geq0\right\} $
. Hence $E[X(1)]=0$ and $Var[X(1)]$=0. We denote the characteristic
function of $X(1)$ (the mother distribution) by $\phi(u)=E[\exp(iuX(1))]$.
We note that, similar to a Brownian Motion, we have $E[X(t)]=1$ and
$Var[X(t)]$=t and hence $Var[\sigma X(t)]=\sigma^{2}t$.

If we set $\omega$ in \eqref{eq:sdeLevyImpliedVol} to be $\omega=-\log\left(\phi\left(-\sigma i\right)\right)$,
we call the volatility parameter $\sigma$ needed to match the model
price with a given market price the implied Levy space volatility
of the option.

To define the implied Levy time volatility we start from a similar
Levy process $X$ and we consider that the dynamics of the underlying
are given as 
\begin{eqnarray}
S(t) & = & S_{0}\exp\left[\left(r-q+\omega\sigma^{2}\right)t+X(t)\right]\label{eq:sdeLevyImpliedVolTime}\\
\omega & = & \log\left(\phi(-i)\right)\nonumber 
\end{eqnarray}

Given a market price, we now use the terminology of implied Levy time
volatility of the option to describe the volatility parameter $\sigma$
needed to match the model price with the market price. Note that in
the $\bsm$ setting the distribution (and hence also the corresponding
vanilla option prices) of $\sigma W(t)$ and $W(\sigma^{2}t)$ is
the same, namely a $\mathcal{N}(0,\sigma^{2}t)$ distribution, but
this is not necessary the case for the more general Levy cases.

The price of an European option is done using characteristic functions
through the Carr-Madan formula \cite{Carr_Madan_1998} and the procedure
is specialized to various Levy processes: normal inverse Gaussian
(NIG), Meixner, etc.

\section{Volatility surface based on models for the dynamics of implied volatility}

In the literature there are two distinct directions for treatment
and construction of volatility surfaces \cite{Carr_Wu_2010}. One
approach assumes dynamics for the underlying that can accommodate
the observed implied volatility smiles and skews. As we have seen
in previous chapters, such approaches include stochastic volatility
models as well as various Levy processes. The general procedure is
to estimate the coefficients of the dynamics through calibration from
observed option prices. Another approach relies on explicitly specifying
dynamics of the implied volatilities, with models belonging to this
class being termed {}``market models'' of implied volatility. In
general, this approach assumes that the entire implied volatility
surface has known initial value and evolves continuously over time.
The approach fi{}rst specifi{}es the continuous martingale component
of the volatility surface, and then derives the restriction on the
risk-neutral drift of the surface imposed by the requirement that
all discounted asset prices be martingales. Such models are presented
in \cite{Avellaneda_zhu_1998,Fengler_2005,Hafner_2004,Daglish_et_al_2007}
An approach that was described as falling between the two categories
was described in \cite{Carr_Wu_2010} and is described next

\subsection{Carr and Wu approach}

Similar to the market model approach, it directly models the dynamics
of implied volatilities. However, instead of taking the initial implied
volatility surface as given and infer the risk-neutral drift of the
implied volatility dynamics, both the risk-neutral drift and the martingale
component of the implied volatility dynamics are specified, from which
the allowable shape for the implied volatility surface is derived.
The shape of the initial implied volatility surface is guaranteed
to be consistent with the specifi{}ed implied volatility dynamics
and, in this sense, this approach is similar to the fi{}rst category.

The starting point is the assumption that a single standard Brownian
motion drives the whole volatility surface, and that a second partially
correlated standard Brownian motion drives the underlying security
price dynamics. By enforcing the condition that the discounted prices
of options and their underlying are martingales under the risk-neutral
measure, one obtains a partial differential equation (PDE) that governs
the shape of the implied volatility surface, termed as Vega-Gamma-Vanna-Volga
(VGVV) methodology, since it links the theta of the options and their
four Greeks. Plugging in the analytical solutions for the $\bsm$
Greeks, the PDE is reduced into an algebraic relation that links the
shape of the implied volatility surface to its risk-neutral dynamics.

By parameterizing the implied variance dynamics as a mean-reverting
square-root process, the algebraic equation simplifi{}es into a quadratic
equation of the implied volatility surface as a function of a standardized
moneyness measure and time to maturity. The coeffi{}cients of the
quadratic equation are governed by six coeffi{}cients related to the
dynamics of the stock price and implied variance. This model is denoted
as the square root variance (SRV) model. 

Alternatively, if the implied variance dynamics is parametrized as
a mean-reverting lognormal process, one obtains another quadratic
equation of the implied variance as a function of log strike over
spot and time to maturity. The whole implied variance surface is again
determined by six coeffi{}cients related to the stock price and implied
variance dynamics. This model is labeled as the lognormal variance
(LNV) model.

The computational cost for calibration is quite small, since computing
implied volatilities from each of the two models (SRV and LNV) is
essentially based on solving a corresponding quadratic equation.

The calibration is based on setting up a state-space framework by
considering the model coeffi{}cients as hidden states and regarding
the fi{}nite number of implied volatility observations as noisy observations.
The coeffi{}cients are inferred from the observations using an unscented
Kalman fi{}lter.

Let us introduce the framework now. We note that zero rates are assumed
without loss of generality.

The dynamics of the stock price of the underlying are assumed to be
given by 
\[
dS(t)=S(t)\sqrt{v(t)}dW(t)
\]

with dynamics of the instantaneous return variance $v(t)$ left unspecified. 

For each option struck at $K$ and expiring at $T$, its implied volatility
$I(t;K,T)$ follows a continuous process given by
\[
dI(t;K,T)=\mu(t)dt+\omega(t)dZ(t)
\]

where $Z(t)$ is a Brownian motion.

The drift $\mu(t)$ and volatility of volatility $\omega(t)$ can
depend on $K,T$ and $I(t;K,T)$

We also assume that we have the following correlation relationship
\[
\rho(t)dt=E\left[dW(t)dZ(t)\right]
\]

The relationship $I(t;K,T)>0$ guarantees that there is no static
arbitrage between any option at $(K;T)$ and the underlying stock
and cash.

It is further required that no dynamic arbitrage (NDA) be allowed
between any option at $(K;T)$ and respectively a basis option at
$(K_{0};T_{0})$ and the stock.

For concreteness, let the basis option be a call with $C(t;T,K)$
denoting its value, and let all other options be puts, with $P(t;K,T)$
denoting the corresponding values. We can write both the basis call
and other put options in terms of the $\bsm$ put formula:
\begin{eqnarray*}
P(t;K,T) & = & BSM\left(S(t),I\left(t;K,T\right),t\right)\\
C(t;K_{0},T_{0}) & = & BSM\left(S(t),I\left(t;K_{0},T_{0}\right),t\right)+S(t)-K_{0}
\end{eqnarray*}

We can form a portfolio between the two to neutralize the exposure
on the volatility risk $dZ$
\[
\part{BSM}{\sigma}\left(S(t),I\left(t;K,T\right),t\right)\omega(K,T)-N^{c}(t)\part{BSM}{\sigma}\left(S(t),I\left(t;K_{0},T_{0}\right),t\right)\omega(K_{0},T_{0})=0
\]

We can further use $N^{S}(t)$ shares of the underlying stock to achieve
delta neutrality:
\[
BSM\left(S(t),I\left(t;K,T\right),t\right)-N^{c}(t)\left[1+BSM\left(S(t),I\left(t;K_{0},T_{0}\right),t\right)\right]-N^{S}(t)=0
\]

Since shares have no Vega, this three-asset portfolio retains zero
exposure to $dZ$ and by construction has zero exposure to $dW$.

By Ito's lemma, each option in this portfolio has risk-neutral drift
(RND) given by
\[
RND=\part{BSM}t+\mu(t)\part{BSM}{\sigma}+\frac{v(t)}{2}S^{2}(t)\part{^{2}BSM}{S^{2}}+\rho(t)\omega(t)\sqrt{v(t)}S(t)\part{^{2}BSM}{\sigma\partial S}+\frac{\omega^{2}(t)}{2}\part{^{2}BSM}{\sigma^{2}}
\]

Note: For simplicity of notations, for the remainder of the chapter
we will use $B$ instead of $BSM$

No dynamic arbitrage and no rates imply that both option drifts must
vanish, leading to the fundamental {}``PDE\textquotedbl{}. 
\begin{equation}
-\part Bt=\mu(t)\part B{\sigma}+\frac{v(t)}{2}S^{2}(t)\part{^{2}B}{S^{2}}+\rho(t)\omega(t)\sqrt{v(t)}S(t)\part{^{2}B}{\sigma\partial S}+\frac{\omega^{2}(t)}{2}\part{^{2}B}{\sigma^{2}}\label{eq:fundPDE}
\end{equation}

The class of implied volatility surfaces defined by the fundamental
{}``PDE\textquotedbl{} \eqref{eq:fundPDE} is termed the Vega-Gamma-Vanna-Volga
(VGVV) model

We should note that \eqref{eq:fundPDE} is not a PDE in the traditional
sense for the following reasons
\begin{itemize}
\item Traditionally, a PDE is specified to solve the value function. In
our case, the value function $B\left(S(t),I(t;K,T),t\right)$ is well-known.
\item The coefficients are deterministic in traditional PDEs, but are stochastic
in \eqref{eq:fundPDE}
\end{itemize}
The {}``PDE'' is not derived to solve the value function, but rather
it is used to show that the various stochastic quantities have to
satisfy this particular relation to exclude dynamic arbitrage. Plugging
in the $\bsm$ formula for $B$ and its partial derivatives $\part Bt,\part{^{2}B}{S^{2}},\part{^{2}B}{S\partial\sigma},\part{^{2}B}{\sigma^{2}}$,
we can reduce the {}``PDE\textquotedbl{} constraint into an algebraic
restriction on the shape of the implied volatility surface $I(t;K,T)$
\[
\frac{I^{2}(t;K,T)}{2}-\mu(t)I(t;K,T)\tau-\left[\frac{v(t)}{2}-\rho(t)\omega(t)\sqrt{v(t)}\sqrt{\tau}d_{2}+\frac{\omega^{2}(t)}{2}d_{1}d_{2}\tau\right]=0
\]

where $\tau=T-t$

This algebraic restriction is the basis for the specific VGVV models:
SRV and LNV, that we describe next.

For SRV we assume square-root implied variance dynamics
\[
dI^{2}(t)=\kappa(t)\left[\theta(t)-I^{2}(t)\right]dt+2w(t)e^{-\eta(t)\left(T-t\right)}I(t)dZ(t)
\]

If we represent the implied volatility surface in terms of $\tau=T-t$
and standardized moneyness $z(t)\triangleq\frac{\ln\left(\nicefrac{K}{S(t)}\right)+0.5I^{2}\tau}{I\sqrt{\tau}}$,
then $I(z,\tau)$ solves the following quadratic equation
\begin{eqnarray*}
\left(1+\kappa(t)\right)I^{2}(z,\tau)+\left(w^{2}(t)e^{-2\eta(t)\tau}\tau^{1.5}z\right)I(z,\tau)\\
-\left[\left(\kappa(t)\theta(t)-w^{2}(t)e^{-2\eta(t)\tau}\right)\tau+v(t)+2\rho(t)\sqrt{v(t)}e^{-\eta(t)\tau}\sqrt{\tau}z+w^{2}(t)e^{-2\eta(t)\tau}\tau z^{2}\right] & = & 0
\end{eqnarray*}

For LNV we assume log-normal implied variance dynamics
\[
dI^{2}(t)=\kappa(t)\left[\theta(t)-I^{2}(t)\right]dt+2w(t)e^{-\eta(t)\left(T-t\right)}I(t)dZ(t)
\]

If we represent the implied volatility surface in terms of $\tau=T-t$
and log relative strike $k(t)\triangleq\ln\left(\nicefrac{K}{S(t)}\right)$,
then $I(k,\tau)$ solves the following quadratic equation
\begin{eqnarray*}
\frac{w^{2}(t)}{4}e^{-2\eta(t)\tau}\tau^{2}I^{4}(k,\tau)+\left[1+\kappa(t)\tau+w^{2}(t)e^{-2\eta(t)\tau}\tau-\rho(t)\sqrt{v(t)}w(t)e^{-\eta(t)\tau}\right]I^{2}(k,\tau)\\
-\left[v(t)+\kappa\theta(t)\tau+2\rho(t)\sqrt{v(t)}e^{-\eta(t)\tau}k+w^{2}(t)e^{-2\eta(t)\tau}k^{2}\right] & = & 0
\end{eqnarray*}

For both SRV and LNV models we have six time varying stochastic coefficients:
\[
\kappa(t),\theta(t),w(t),\eta(t),\rho(t),v(t)
\]

Given time $t$ values for the six coefficients, the whole implied
volatility surface at time $t$ can be found as solution to quadratic
equations.

The dynamic calibration procedure treats the six coefficients as a
state vector $X(t)$ and it assumes that $X(t)$ propagates like a
random walk
\[
X(t)=X(t-1)+\sqrt{\Sigma_{X}}\epsilon(t)
\]

where $\Sigma_{X}$ is a diagonal matrix. It also assumes that all
implied volatilities are observed with errors $IID$ normally distributed
with error variance $\sigma_{e}^{2}$ 
\[
y(t)=h(X(t))+\sqrt{\Sigma_{y}}\epsilon(t)
\]

with $h(\cdot)$ denoting the model value (quadratic solution for
SRV or LNV) and $\Sigma_{y}=I_{N}\sigma_{e}^{2}$, with $I_{N}$ denoting
an identity matrix of dimension $N$

This setup introduces seven auxiliary parameters $\Theta$ that defi{}ne
the covariance matrix of the state and the measurement errors.

When the state propagation and the measurement equation are Gaussian
linear, the Kalman fi{}lter provides effi{}cient forecasts and updates
on the mean and covariance of the state and observations. The state-propagation
equations are Gaussian and linear, but the measurement functions $h\left(X(t)\right)$
are not linear in the state vector. To handle the non-linearity we
employ the unscented Kalman fi{}lter. For additional details the reader
is referred to \cite{Carr_Wu_2010}.

The procedure was applied successfully on both currency options and
equity index options, and compared with Heston.

The comparison with Heston provided the following conclusions:
\begin{itemize}
\item generated half the root mean squared error
\item explains 4\% more variation
\item generated errors with lower serial correlation
\item can be calibrated 100 times faster
\item The whole sample (573 weeks) of implied volatility surfaces can be
fitted in about half a second (versus about 1 minute for Heston).
\end{itemize}

\section{Volatility surface based on parametric representations}

Various parametric or semi-parametric representations of the volatility
surface have been considered in the literature. A recent overview
was given in \cite{Fengler_2010}.

\subsection{Polynomial parametrization }

A popular representation was suggested in \cite{Dumas_et_al_1998},
which proposed that the implied volatility surface is modeled as a
quadratic function of the moneyness $\mathcal{M}\triangleq\nicefrac{\ln\left(\nicefrac{F}{K}\right)}{\sqrt{T}}$
\[
\sigma\left(\mathcal{M},T\right)=b_{1}+b_{2}\mathcal{M}+b_{3}\mathcal{M}^{2}+b_{4}T+b_{5}\mathcal{M}T
\]

This model was considered for oil markets in \cite{Borovkova_Permana_2009},
concluding that the model gives only an \textquotedblleft{}average\textquotedblright{}
shape, due to its inherent property of assuming the quadratic function
of volatility versus moneyness to be the same across all maturities.
Note that increasing the power of the polynomial volatility function
(from two to three or higher) does not really offer a solution here,
since this volatility function will still be the same for all maturities. 

To overcome those problems a semi parametric representation was considered
in \cite{Borovkova_Permana_2009}, where they kept quadratic parametrization
of the volatility function for each maturity $T$, and approximate
the implied volatility by a quadratic function which has time dependent
coefficients. 

A similar parametrization (but dependent on strike and not moneyness)
was considered in \cite{Christoffersen_Jacobs_2004} under the name
Practitioner\textquoteright{}s BlackScholes. It was shown that outperforms
some other models in terms of pricing error in sample and out of sample. 

Such parametrizations may some certain drawbacks, such as: 
\begin{itemize}
\item are not designed to ensure arbitrage-free of the resulting volatility
surface 
\item the dynamics of the implied volatility surface may not be adequately
captured 
\end{itemize}
We now describe other parametrizations that may be more suitable.

\subsection{Stochastic volatility inspired (SVI) parametrization }

SVI is a practitioner designed parametrization \cite{Gatheral_2004,Gatheral_2006}.
Very recent papers provide the theoretical framework and describe
its applicability to energy markets \cite{Deryabin_2010,Deryabin_2011}.
We also note that SVI procedure may be employed together with conditions
for no vertical and horizontal spread arbitrage, such as in \cite{Gurrieri_2010}.

The essence of SVI is that each time slice of the implied volatility
surface is fitted separately, such that in the logarithmic coordinates
the implied variance curve is a hyperbola, and additional constraints
are imposed that ensure no vertical/ horizontal spread arbitrage opportunities.
The hyperbola is chosen because it gives the correct asymptotic representation
of the variance when log-strike tends to plus or minus infinity: written
as a function of $\ln\left(\nicefrac{K}{F}\right)$, where $K$ is
the strike and $F$ is the forward price, and time being fixed, the
variance tends asymptotically to straight lines when $\ln\left(\nicefrac{K}{F}\right)\rightarrow\pm\infty$

The parametrization form is on the implied variance: 
\[
\sigma^{2}\left[x\right]\triangleq v(\left\{ m,s,a,b,\rho\right\} ,x)=a+b\left(\rho\left(x-m\right)+\sqrt{\left(kx-m\right)^{2}+s^{2}}\right)
\]

where $a,b,\rho,m,s$ are parameters which are dependent on the time
slice and $x=\ln\left(\nicefrac{K}{F}\right)$.

We should note that it was recently shown \cite{Roper_2010} that
SVI may not be arbitrage-free in all situations. Nevertheless SVI
has many advantages such as small computational time, relatively good
approximation for implied volatilities for strikes deep in- and out-of-the-money.
The SVI fit for equity markets is much better than for energy markets,
for which \cite{Deryabin_2011} reported an error of maximum 4-5\%
for front year and respectively 1-2\% for long maturities. 

Quasi explicit calibration of SVI is presented in \cite{DeMarco_2009},
based on dimension reduction for the optimization problem. The original
calibration procedure is based on matching input market data $\left\{ \sigma_{i}^{MKT}\right\} _{i=1...M}$,
which becomes an optimization problem with five variables: $a,b,\rho,m,s$:
\[
\min_{\left\{ a,b,\rho,m,s\right\} }\sum_{i=1}^{N}\left(v\left[\left\{ m,s,a,b,\rho\right\} ,\ln\left(\frac{K_{i}}{F}\right)\right]-\left(\sigma_{i}^{MKT}\right)^{2}\right)^{2}
\]

The new procedure is based on a change of variables
\[
y=\frac{x-m}{s}
\]

Focusing on total variance $V=vT$, the SVI parametrization becomes
\[
V(y)=\alpha T+\delta y+\beta\sqrt{y^{2}+1}
\]

where we have used the following notations
\begin{eqnarray*}
\beta & = & bsT\\
\delta & = & \rho bsT\\
\alpha & = & aT
\end{eqnarray*}

We also use the notation $\Vbsq i=\mktvolsq iT$

Therefore, for given $m$ and $s$, which is transformed into $\left\{ y_{i},\bar{V}_{i}\right\} $,
we look for the solution of the 3-dimensional problem
\begin{equation}
\min_{\{\beta,\delta,\alpha\}}F_{\{y_{i},\bar{V}_{i}\}}(\b,\d,\a)\label{eq:SVIthree}
\end{equation}

with the objective functional for reduced dimensionality problem defined
by
\[
F_{\{y_{i},v_{i}\}}(\b,\d,\a)=\sum_{i=1}^{N}w_{i}\left(\a+\d y_{i}+\beta\sqrt{y_{i}^{2}+1}-\Vbsq i\right)^{2}
\]

The domain on which to solve the problem is defined as
\[
\begin{cases}
\begin{array}{c}
\beta_{MIN}\leq\beta\leq4s\\
-\beta\leq\delta\leq\beta\\
-\left(4s-\beta\right)\leq\delta\leq\left(4s-\beta\right)\\
\alpha_{MIN}\leq\alpha\leq\Vbmax
\end{array}\end{cases}
\]

For a solution $\left\{ \b^{*},\d^{*},\a^{*}\right\} $of the problem
\eqref{eq:SVIthree}, we identify the corresponding triplet $\left\{ a^{*},b^{*},\rho^{*}\right\} $
and then we solve the 2-dimensional optimization problem 
\[
\min_{\left\{ m,s\right\} }\sum_{i=1}^{N}\left(v\left[\left\{ m,s,a^{*},b^{*},\rho^{*}\right\} ,\ln\left(\frac{K_{i}}{F}\right)\right]-v_{i}^{MKT}\right)^{2}
\]

Thus the original calibration problem was cast as a combination of
distinct 2-parameter optimization problem and, respectively, 3-parameter
optimization problem. Because the {}``2+3'' procedure is much less
sensitive to the choice of initial guess, the resulting parameter
set is more reliable and stable. For additional details the reader
is referred to \cite{DeMarco_2009}. The SVI parametrization is performed
sequentially, expiry by expiry. An enhanced procedure was presented
in \cite{Gurrieri_2010} to obtain a satisfactory term structure for
SVI, which satisfies the no-calendar spread arbitrage in time while
preserving the condition of no-strike arbitrage.

\subsection{Entropy based parametrization }

Entropic calibrations have been considered by a number of authors.
It was done for risk-neutral terminal price distribution, implied
volatility function and the option pricing function.

An algorithm that yields an arbitrage-free diffusion process by minimizing
the relative entropy distance to a prior diffusion is described in
\cite{Avellaneda_et_al_1997}. This results in an efficient calibration
method that can match an arbitrary number of option prices to any
desired degree of accuracy. The algorithm can be used to interpolate,
both in strike and expiration date, between implied volatilities of
traded options. 

Entropy maximization is employed in \cite{Buchen_Kelly_1997} and
\cite{Brody_et_al_2007} to construct the implied risk-neutral probability
density function for the terminal price of the underlying asset. The
advantage of such an entropic pricing method is that it does not rely
on the use of superfluous parameters, and thus avoids the issue of
over fitting altogether. Furthermore, the methodology is flexible
and universal in the sense that it can be applied to a wider range
of calibration situations. 

Most of the entropy-based calibration methodologies adopted in financial
modeling, whether they are used for relative entropy minimization
or for absolute entropy maximization, rely on the use of the logarithmic
entropy measure of Shannon and Wiener. One drawback in the use of
logarithmic entropy measures is that if the only source of information
used to maximize entropy is the market prices of the vanilla options,
then the resulting density function is necessarily of exponential
form. On the other hand, empirical studies indicate that the tail
distributions of asset prices obey power laws of the Zipf\textendash{}Mandelbrot
type \cite{Brody_et_al_2007} . Thus we would like to employ entropies
that may recover power law distribution, such as Renyi entropy \cite{Brody_et_al_2007} 

Maximization of Renyi entropy is employed to obtain arbitrage-free
interpolation. The underlying theoretical idea is that, irrespective
of the nature of the underlying price process, the gamma associated
with a European-style vanilla option always defines a probability
density function of the spot price implied by the existence of the
prices for option contracts. There is a one-to-one correspondence
between the pricing formula for vanilla options and the associated
gamma. Therefore, given option gamma we can unambiguously recover
the corresponding option pricing formula. 

We present here an overview of the approach presented in \cite{Brody_et_al_2007}

Given strikes $K_{j},\, j=1...M$, corresponding for input market
prices, maximizing the Renyi entropy yields a density function of
the form: 
\begin{equation}
p(x)=\left(\lambda+\beta_{0}x+\sum_{j=1}^{M}\beta_{j}\left(x-K_{j}\right)^{+}\right)^{\frac{1}{\alpha-1}}\label{eq:densityRenyi}
\end{equation}

The parameters $\alpha,\lambda,\beta_{0},...,\beta_{M}$ are calibrated
by matching the input prices to the prices computed using the density
function \eqref{eq:densityRenyi}. 

We exemplify for the case of call options. For each $j=1...M$, we
have to impose the matching condition to market price $C_{j}^{MKT}$
\[
\bar{S}_{0}-K_{m}-\frac{\alpha-1}{\alpha}\sum_{m=1}^{j-1}Y_{m}\left[X_{m}\left(x\right)\right]^{\frac{\alpha}{\alpha-1}}\left(x-K_{m}-\frac{\alpha-1}{2\alpha-1}Y_{m}X_{m}\left(x\right)\right)|_{x=K_{m}}^{x=K_{m+1}}=C_{j}^{MKT}
\]

where 
\begin{eqnarray*}
X_{j}(x) & \triangleq & \lambda+\sum_{j=1}^{M}\beta_{j}\left(x-K_{j}\right)\\
Y_{j} & \triangleq & \left(\sum_{m=0}^{j}\beta_{m}\right)
\end{eqnarray*}

We also impose the normalization condition
\[
\intop_{0}^{\infty}p(x)dx=1\Longrightarrow\frac{\alpha-1}{\alpha}\sum_{j=0}^{M}Y_{j}\left[X_{j}\left(x\right)\right]^{\frac{\alpha}{\alpha-1}}|_{x=K_{j}}^{x=K_{j+1}}=1
\]

Since the density function is explicitly given, is straightforward
to use for calibration additional option types, such as digitals or
variance swaps. 

The result is described in \cite{Brody_et_al_2007} as leading to
the power-law distributions often observed in the market. By construction,
the input data are calibrated with a minimum number of parameters,
in an efficient manner. The procedure allows for accurate recovery
of tail distribution of the underlying asset implied by the prices
of the derivatives. One disadvantage is that the input values are
supposed to be arbitrage free, otherwise the algorithm will fail.
It is possible to enhance the algorithm to handle inputs with arbitrage,
but the resulting algorithm will lose some of the highly efficient
characteristics, since now we need to solve systems of equations in
a least square sense

\subsection{Parametrization using weighted shifted lognormal distributions}

A weighted sum of interpolation functions taken in a parametric family
is considered in the practitioner papers \cite{Bloch_2010,Bloch_Coello_Coello_2010}
to generate a surface without arbitrage in time and in space, while
remaining as closely as possible to market data. Each function in
the family is required to satisfy the no-free lunch constraints, specified
later, in such way that they are preserved in the weighted sum. 

In this parametric model, the price of a vanilla option price of strike
$K$ and maturity $T$ is estimated at time $t_{0}=0$ by the weighted
sum of N functionals 
\[
\sum_{i=1}^{N}a_{i}(T)F_{i}\left(t_{0,}S_{0},P(T);K,T\right)
\]

with $a_{i}(T)$ weights and $P(t)=B\left(0,t\right)$ the zero coupon
bond price. 

Several families $F_{i}$ can satisfy the No-Free-Lunch constraints,
for instance a sum of lognormal distributions, but in order to match
a wide variety of volatility surfaces the model has to produce prices
that lead to risk-neutral pdf of the asset prices with a pronounced
skew. If all the densities are centered in the log-space around the
forward value, one recovers the no-arbitrage forward pricing condition
but the resulting pricing density will not display skew. However,
centering the different normal densities around different locations
(found appropriately) and constraining the weights to be positive,
we can recover the skew. Since we can always convert a density into
call prices, we can then convert a mixture of normal densities into
a linear combination of $\bsm$ formulae. 

Therefore, we can achieve that goal with a sum of shifted log-normal
distributions, that is, using the $\bsm$ formula with shifted strike
(modified by the parameters) as an interpolation function
\[
F_{i}\left(t_{0},S_{0},P(T);K,T\right)=Call_{BSM}\left(t_{0},S_{0},P(T),\hat{K}\left(1+\mu_{i}(T)\right),T,\sigma_{i}\right)
\]

with $\hat{K}$denoting adjusted strike.

We note that the value of strike is adjusted only if we apply the
procedure for equity markets, in which case it becomes
\[
\hat{K}(K,t)=K+D(0,t)
\]

with $D(0,t)$ is the compounded sum of discrete dividends between
$0$ and $t$.

The no-arbitrage theory imposes time and space constraints on market
prices. Hence, the time dependent parameters $a_{i}(t)$ and $\mu_{i}(t)$
are used to recover the time structure of the volatility surface.
It is argued that it sufficient to use a parsimonious representation
of the form
\begin{eqnarray*}
\mu_{i}(t) & = & \mu_{i}^{0}f\left(t,\beta_{i}\right)\\
a_{i}(t) & = & \left(\sum_{i=1}^{N}\frac{a_{i}^{0}}{f\left(t,\beta_{i}\right)}\right)\frac{a_{i}^{0}}{f\left(t,\beta_{i}\right)}\\
f\left(t,\beta_{i}\right) & \triangleq & 1-\frac{2}{1+\left(1+\frac{t}{\beta}\right)^{2}}
\end{eqnarray*}

Making the weights and the shift parameter time-dependent to fit a
large class of volatility surfaces leads to the following no-free
lunch constraints, for any time $t$
\begin{itemize}
\item $a_{i}\geq0$ to get convexity of the price function
\item $\sum_{i=1}^{N}a_{i}(t)=1$ to get a normalized risk-neutral probability
\item $\sum_{i=1}^{N}a_{i}(t)\mu_{i}(t)=1$ to keep the martingale property
of the induced risk-neutral pdf 
\item $\mu_{i}(t)$ to get non-degenerate functions 
\end{itemize}
The model being invariant when multiplying all the terms $a_{i}^{0}$
with the same factor, we impose the normalization constraint 
\[
\sum_{i=1}^{N}a_{i}^{0}=1
\]
to avoid the possibility of obtaining different parameter sets which
nevertheless yield the same model. 

Given the $N$ parameters and assuming a constant volatility $\sigma_{i}(t)=\sigma_{i}^{0}$,
there are $4N-2$ free parameters for the$N$-function model since
we can use the constraints to express $a_{1}^{0}$ and, respectively,
$a_{1}^{0}\mu_{1}^{0}$ in terms of $\left\{ a_{i}^{0}\right\} _{i=1...N}$
and $\left\{ \mu_{i}^{0}\right\} _{i=1...N}$ (see also Appendix A
of \cite{Bloch_Coello_Coello_2010}).

As such, this model does not allow for the control of the long term
volatility surface. Therefore, for the model to be complete we specify
the time-dependent volatility such that it captures the term structure
of the implied volatility surface: 
\[
\sigma_{i}\left(t\right)=\gamma_{i}e^{-c_{i}t}+d_{i}f\left(t,b_{i}\right)
\]

Thus we need to solve a $7N-2$ optimization problem. This is done
in \cite{Bloch_2010,Bloch_Coello_Coello_2010} using a global optimizer
of Differential Evolution type.

\section{Volatility surface based on nonparametric representations, smoothing
and interpolation }

This broad set of procedures may be divided into several categories.

\subsection{Arbitrage-free algorithms}

Interpolation techniques to recover a globally arbitrage-free call
price function have been suggested in various papers, e.g., \cite{Kahale_2004,Wang_et_al_2004}.
A crucial requirement for these algorithms to work that the data to
be interpolated are arbitrage-free from the beginning. \cite{Kahale_2004}
proposes an interpolation procedure based on piecewise convex polynomials
mimicking the $\bsm$ pricing formula. The resulting estimate of the
call price function is globally arbitrage-free and so is the volatility
smile computed from it. In a second step, the total (implied) variance
is interpolated linearly along strikes. Cubic B-spline interpolation
was employed by \cite{Wang_et_al_2004}, with interpolation performed
on option prices, and the shape restrictions in interpolated curves
was imposed by the use of semi-smooth equations minimizing the distance
between the implied risk neutral density and a prior approximation. 

Instead of smoothing prices, \cite{Benko_et_al_2007} suggests to
directly smooth implied volatility parametrization by means of constrained
local quadratic polynomials. Let us consider that we have M expiries
$\left\{ T_{j}\right\} $ and N strikes $\left\{ x_{i}\right\} $
, while the market data is denoted by $\left\{ \sigma_{i}^{MKT}(T_{j})\right\} $

Two approaches are considered:

\textbullet{} each maturity is treated separately 

\textbullet{} all maturities are included in the cost functional to
minimize 

The first case implies minimization of the following (local) least
squares criterion at each expiry $T_{j},$~j=1...NT 
\[
\min_{\left\{ \a_{0}^{(j)},\a_{1}^{(j)},\a_{2}^{(j)}\right\} }\sum_{i=1}^{N}\left\{ \sigma_{i}^{MKT}(T_{j})-\a_{0}^{(j)}-\a_{1}^{(j)}\left(x_{i}-x\right)-\a_{2}^{(j)}\left(x_{i}-x\right)^{2}\right\} \frac{1}{h}\mathcal{K}\left[\frac{x_{i}-x}{h}\right]
\]

where $\mathcal{K}$ is a kernel function, typically a symmetric density
function with compact support.

One example is the Epanechnikov kernel 
\[
\mathcal{K}\left(u\right)=0.75\left(1-u^{2}\right)\mathbf{1}\left[\left|u\right|\leq1\right]
\]

with $\mathbf{1}(A)$ denoting the indicator function for a set A
and $h$ is the bandwidth which governs the trade-off between bias
and variance.

The optimization problem for the second approach is 
\[
\min_{\left\{ \left\{ \a_{0}^{(j)},\a_{1}^{(j)},\a_{2}^{(j)},\a_{3}^{(j)},\a_{4}^{(j)}\right\} _{j=1..M}\right\} }\sum_{j=1}^{M}\sum_{i=1}^{N}\Psi\left(\left\{ \a_{0}^{(j)},\a_{1}^{(j)},\a_{2}^{(j)},\a_{3}^{(j)},\a_{4}^{(j)}\right\} \right)\frac{1}{h_{X}}\mathcal{K}\left[\frac{x_{i}-x}{h_{X}}\right]\frac{1}{h_{T}}\mathcal{K}\left[\frac{T_{j}-T}{h_{T}}\right]
\]

with defined as
\begin{eqnarray*}
\Psi\left(\left\{ \a_{0}^{(j)},\a_{1}^{(j)},\a_{2}^{(j)},\a_{3}^{(j)},\a_{4}^{(j)}\right\} \right) & \triangleq & \sigma_{i}^{MKT}(T_{j})-\a_{0}^{(j)}-\a_{1}^{(j)}\left(x_{i}-x\right)\\
 &  & -\a_{2}^{(j)}\left(T_{j}-T\right)-\a_{3}^{(j)}\left(x_{i}-x\right)^{2}-\a_{4}^{(j)}\left(x_{i}-x\right)\left(T_{j}-T\right)
\end{eqnarray*}

The approach yields a volatility surface that respects the convexity
conditions, but neglects the conditions on call spreads and the general
price bounds. Therefore the surface may not be fully arbitrage-free.
However, since convexity violations and calendar arbitrage are by
far the most virulent instances of arbitrage in observed implied volatility
data, the surfaces will be acceptable in most cases.

The approach in \cite{Fengler_2010} is based on cubic spline smoothing
of option prices rather than on interpolation. Therefore, the input
data does not have to be arbitrage-free. It employs cubic splines,
with constraints specifically added to the minimization problem in
order to ensure that there is no arbitrage. A potential drawback for
this approach is the fact that the call price function is approximated
by cubic polynomials. This can turn out to be disadvantageous, since
the pricing function is not in the domain of polynomials functions.
It is remedied by the choice of a sufficiently dense grid in the strike
dimension. 

Instead of cubic splines, \cite{Laurini_2007} employs constrained
smoothing B-splines. This approach permits to impose monotonicity
and convexity in the smoothed curve, and also through additional pointwise
constraints. According to the author, the methodology has some apparent
advantages on competing methodologies. It allows to impose directly
the shape restrictions of no-arbitrage in the format of the curve,
and is robust the aberrant observations. Robustness to outliers is
tested by comparing the methodology against smoothing spline, Local
Polynomial Smoothing and Nadaraya-Watson Regression. The result shows
that Smoothing Spline generates an increasing and non-convex curve,
while the Nadaraya-Watson and Local Polynomial approaches are affected
by the more extreme points, generating slightly non convex curves.

It is mentioned in \cite{Orosi_2010} that a large drawback of bi-cubic
spline or B-spline models is that they require the knots to be placed
on a rectangular grid. Correspondingly, it considers instead a thin-spline
representation, allowing arbitrarily placed knots. This leads to a
more complex representation at shorter maturities while preventing
overfitting.

Thin-spline representation of implied volatility surface was also
considered in \cite{Brecher_2005} and section 2.4 of \cite{vanDerKamp_2009},
where it was used to obtain a pre-smoothed surface that will be eventually
used as starting point for building a local volatility surface. 

An efficient procedure was shown in \cite{Maruhn_2010} for constructing
the volatility surface using generic volatility parametrization for
each expiry, with no-arbitrage conditions in space and time being
added as constraints, while a regularization term was added to the
calibrating functional based on the difference between market implied
volatilities and, respectively, volatilities given by parametrization.
Bid-ask spread is also included in the setup. The resulting optimization
problem has a lot of sparsity/structure, characteristics that were
exploited for obtaining a good fit in less than a second

\subsection{Remarks on spline interpolation }

The following splines are usually employed to interpolate implied
volatilities

\textbullet{} Regular cubic splines

\textbullet{} Cubic B-splines 

\textbullet{} Thin splines

Certain criteria (such as arbitrage free etc) have to be met, and
relevant papers were described in the previous section . Here we just
refer to several generic articles on spline interpolation.

\cite{Wolberg_Alfy_2002} describes an approach that yields monotonic
cubic spline interpolation. 

Although somewhat more complicated to implement, B-splines may be
preferred to cubic splines, due to its robustness to bad data, and
ability to preserve monotonicity and convexity. A recent paper \cite{Kong_Ong_2009}
describes a computationally efficient approach for cubic B-splines. 

A possible alternative is the thin-plate spline, which gets its name
from the physical process of bending a thin plate of metal. A thin
plate spline is the natural two-dimensional generalization of the
cubic spline, in that it is the surface of minimal curvature through
a given set of two-dimensional data points.

\subsection{Remarks on interpolation in time}

In some situations we need to perform interpolation in time. While
at a first glance it may seem straightforward, special care has to
be employed to ensure that the result still satisfies practical arbitrage
conditions. For example, one should expect that there is no calendar
spread arbitrage \cite{Carr_Madan_2005,Deryabin_2010,Gatheral_2004,Reiner_2004} 

One common approach is to perform linear interpolation in variance.
A variant of it, denoted \textquotedblleft{}total variance interpolation\textquotedblright{},
is described in \cite{Castagna_2007}.

\subsection{ Interpolation based on fully implicit finite difference discretization
of Dupire forward PDE }

We present an approach described in \cite{Andreasen_Huge_2010,Andreasen_Huge_2011,Huge_2010},
based on fully implicit finite difference discretization of Dupire
forward PDE.

We start from the Dupire forward PDE in time-strike space 
\[
-\part ct+\frac{1}{2}\left[\sigma\left(t,k\right)\right]^{2}\part{^{2}c}{k^{2}}=0
\]

Let us consider that we have the following time grid $0=t_{0}<t_{1}<...<t_{N}$
and define $\triangle t_{i}\triangleq t_{i+1}-t_{i}$ 

A discrete (in time) version of the forward equation is 
\[
\frac{c\left(t_{i+1},k\right)-c\left(t_{i},k\right)}{\triangle t_{i}}=\frac{1}{2}\left[\sigma\left(t_{i},k\right)\right]^{2}\part{^{2}c}{k^{2}}\left(t_{i+1},k\right)
\]

This is similar to an implicit finite difference step. It can be rewritten
as 
\begin{equation}
\left[1-\frac{1}{2}\triangle t_{i}\left[\sigma\left(t_{i},k\right)\right]^{2}\part{^{2}}{k^{2}}\right]c\left(t_{i+1},k\right)=c\left(t_{i},k\right)\label{eq:similarImplicit}
\end{equation}

Let us consider that the volatility function is piecewise defined
on the time interval $t_{i}\leq t<t_{i+1}$ and we denote by $\nu_{i}(k)$
the corresponding functions
\[
\nu_{i}(k)\triangleq\sigma(t,k)\,\,\,\,\,\,\, for\,\, t_{i}\leq t<t_{i+1}
\]

Using \eqref{eq:similarImplicit} we can construct European (call)
option prices for all discrete time points for a given a set of volatility
functions $\left\{ \nu_{i}(k)\right\} _{i=1...N}$ by recursively
solving the forward system
\begin{eqnarray}
\left[1-\frac{1}{2}\triangle t_{i}\left[\sigma\left(t_{i},k\right)\right]^{2}\part{^{2}}{k^{2}}\right]c\left(t_{i+1},k\right) & = & c\left(t_{i},k\right)\label{eq:recursiveFwdsystem}\\
c(0,k) & = & \left[S(0)-k\right]^{+}\nonumber 
\end{eqnarray}

Let us discretize the strike space as $K_{MIN}=k_{0}<k_{1}<...<k_{M}=k_{MAX}$

By replacing the differential operator $\nicefrac{\partial^{2}}{\partial k^{2}}$
by the central difference operator 
\begin{eqnarray*}
\delta_{kk}f(k) & = & \frac{2}{\left(k_{j}-k_{j-1}\right)\left(k_{j+1}-k_{j-1}\right)}f(k_{j-1})-\frac{2}{\left(k_{j}-k_{j-1}\right)\left(k_{j+1}-k_{j}\right)}f(k_{j})\\
 &  & +\frac{2}{\left(k_{j+1}-k_{j}\right)\left(k_{j+1}-k_{j-1}\right)}f(k_{j+1})
\end{eqnarray*}

we get the following finite difference scheme system
\begin{eqnarray}
\left[1-\frac{1}{2}\triangle t_{i}\left[\nu_{i}\left(k\right)\right]^{2}\delta_{kk}\right]c\left(t_{i+1},k\right) & = & c\left(t_{i},k\right)\label{eq:systemFD}\\
c(0,k) & = & \left[S(0)-k\right]^{+}\nonumber 
\end{eqnarray}

The matrix of the system \eqref{eq:systemFD} is tridiagonal and shown
in \cite{Andreasen_Huge_2010} to be diagonally dominant, which allows
for a well behaved matrix that can be solved efficiently using Thomas
algorithm \cite{Strikwerda_2004}.Thus we can directly obtain the
European option prices if we know the expressions for $\left\{ \nu_{i}(k)\right\} _{i=1...N}$. 

This suggests that we can use a bootstrapping procedure, considering
that the volatility functions are defined as piecewise constant

Let us first introduce the notations for market data. We consider
that we have a set of discrete option quotes $\left\{ c^{MKT}\left(t_{i},K_{i,p}\right)\right\} $,
where $\left\{ t_{i}\right\} $ are the expiries and $\left\{ K_{i,p}\right\} _{p=1...NK(i)}$
is the set of strikes for expiry $t_{i}$. 

We should note that we may have different strikes for different expiries,
and that $\left\{ K_{i,p}\right\} _{p=1...NK(i)}$ and, respectively,
$\left\{ k_{j}\right\} $ represent different quantities 

Then the piecewise constant volatility functions are denoted as
\[
\nu_{i}(k)\triangleq\begin{cases}
...\\
\sigma_{i,p} & \,\, for\,\, K_{i,p}\leq k<K_{i,p+1}\\
...
\end{cases}
\]

Thus the algorithm consists of solving an optimization problem at
each expiry time, namely
\begin{equation}
\min_{\left\{ a_{i,1},...,a_{i,NK(i)}\right\} }\left[\sum_{p=1}^{NK(i)}\left(c\left(t_{i},K_{i,p}\right)-c^{MKT}\left(t_{i},K_{i,p}\right)\right)^{2}\right]\label{eq:optFwdDupire}
\end{equation}

We remark that, when solving \eqref{eq:optFwdDupire} by some optimization
procedure, one needs to solve only one tridiagonal matrix system for
each optimization iteration.

Regarding interpolation in time, two approaches are proposed in \cite{Andreasen_Huge_2010}.
The first one is based on the formula 
\begin{equation}
\left[1-\frac{1}{2}\left(t-t_{i}\right)\left[\nu_{i}\left(k\right)\right]^{2}\part{^{2}}{k^{2}}\right]c\left(t_{i+1},k\right)=c\left(t_{i},k\right)\,\,\, for\,\, t_{i}<t<t_{i+1}\label{eq:firstInterp}
\end{equation}

while the second one is a generalization of \eqref{eq:firstInterp}
\begin{equation}
\left[1-\frac{1}{2}\left(T(t)-t_{i}\right)\left[\nu_{i}\left(k\right)\right]^{2}\part{^{2}}{k^{2}}\right]c\left(t_{i+1},k\right)=c\left(t_{i},k\right)\,\,\, for\,\, t_{i}<t<t_{i+1}\label{eq:secondInterp}
\end{equation}

where $T(t)$ is a function that satisfies the conditions $T(t_{i})=t_{i}$
and $T'(t)<0$

It is shown in \cite{Andreasen_Huge_2010} that option prices generated
by \eqref{eq:recursiveFwdsystem}and \eqref{eq:firstInterp} and,
respectively, by \eqref{eq:recursiveFwdsystem} and \eqref{eq:secondInterp}are
consistent with the absence of arbitrage in the sense that , for any
pair $\left(t,k\right)$ we have
\begin{eqnarray*}
\part ct\left(t,k\right) & \geq & 0\\
\part{^{2}c}{k^{2}}\left(t,k\right) & \geq & 0
\end{eqnarray*}

\section{Adjusting inputs to avoid arbitrage }

Various papers have tackled the problem of finding conditions that
may be necessary/and or sufficient to ensure that prices/vols are
free of arbitrage \cite{Carr_Madan_2005,Cousot_2007,Davis_Hobson_2007,Herzel_2005,Mercurio,Reiswich_2010}.
If one wants to adjust the set of input prices/vols to avoid arbitrage,
several approaches have been described in the literature. For example,
\cite{aitSahalia_Duarte_2003} presents a relatively simple method
to adjust implied volatilities, such that the resulting set is both
arbitrage free and also closest to the initial values (in a least-squares
sense). Another algorithm is presented in section 8.3 of \cite{Buehler_2008}.
We present in detail the algorithm from \cite{Carr_Madan_2005}, based
on the observation that the absence of call spread, butterfly spread
and calendar spread arbitrages is sufficient to exclude all static
arbitrages from a set of option price quotes across strikes and maturities.

\subsection{Carr and Madan algorithm }

The main idea is as follows: given input market prices and corresponding
bid ask spreads, we start from the price corresponding to first expiry
ATM and adjust the prices for that expiry. We continue to the next
expiry and we make sure that arbitrage constraints are satisfied both
in time and strike space, while adjusting within the bid ask spread. 

We present first the arbitrage constraints from \cite{Carr_Madan_2005},
using notations from there. Let $C_{ij}$ denote the given quote for
a call of strike $K_{i}$ and maturity $T_{j}$. We suppose that the
N strikes $\left\{ K_{i}\right\} $ form an increasing and positive
sequence as do the M maturities $\left\{ T_{j}\right\} $. Without
any loss of generality, we suppose that interest rates and dividends
are zero over the period ending at the longest maturity. 

We augment the provided call quotes with quotes for calls of strike
$K_{0}=0$. For each maturity, these additional quotes are taken to
be equal to the current spot price $S_{0}$. We also take the prices
at maturity $T_{0}=0$ to be $\left(S_{0}-K_{i}\right)^{+}$. This
gives us the augmented matrix of prices $C_{ij}$, with indices $i=0..N$
and $j=1...M$ . 

For each $j>0$ we define the following quantities: 
\begin{eqnarray*}
Q_{i,j} & = & \frac{C_{i-1,j}-C_{i,j}}{K_{i}-K_{i-1}}\\
Q_{0,j} & = & 0
\end{eqnarray*}

For each $i>0$, $Q_{i,j}$ is the cost of a vertical spread which
by definition is long $\nicefrac{1}{\left(K_{i}-K_{i-1}\right)}$
calls of strike $K_{i-1}$ and short $\nicefrac{1}{\left(K_{i}-K_{i-1}\right)}$
calls of strike $K_{i}$. A graph of the payoff from this position
against the terminal stock price indicates that this payoff is bounded
below by zero and above by one. 

We therefore require for our first test that 
\begin{equation}
0\leq Q_{i,j}\leq1,\,\,\, i=1...N,\, j=1...M\label{eq:firstTest}
\end{equation}

Next, for each $j>0$, we define the following quantities: 
\[
BSpr_{i,j}\triangleq C_{i-1,j}-\frac{K_{i+1}-K_{i-1}}{K_{i+1}-K_{i}}C_{i,j}+\frac{K_{i}-K_{i-1}}{K_{i+1}-K_{i}}C_{i+1,j}\,\,\,\,\, i>0
\]

For each $i>0$ $BSpr_{i,j}$ is the cost of a butterfly spread which
by definition is long the call struck at $K_{i-1}$, short $\nicefrac{\left(K_{i+1}-K_{i-1}\right)}{\left(K_{i+1}-K_{i}\right)}$
calls struck at $K_{i}$, and long $\nicefrac{\left(K_{i}-K_{i-1}\right)}{\left(K_{i+1}-K_{i}\right)}$
calls struck at $K_{i+1}$. A graph indicates that the butterfly spread
payoff is non-negative and hence our second test requires that
\[
C_{i-1,j}-\frac{K_{i+1}-K_{i-1}}{K_{i+1}-K_{i}}C_{i,j}+\frac{K_{i}-K_{i-1}}{K_{i+1}-K_{i}}C_{i+1,j}\geq0
\]

Equivalently, we require that
\begin{equation}
C_{i-1,j}-C_{i,j}\geq\frac{K_{i}-K_{i-1}}{K_{i+1}-K_{i}}\left(C_{i,j}-C_{i+1,j}\right)\label{eq:secondTest}
\end{equation}

We define 
\[
q_{i,j}\triangleq Q_{i,j}-Q_{i+1,j}=\frac{C_{i-1,j}-C_{i,j}}{K_{i}-K_{i-1}}-\frac{C_{i,j}-C_{i+1,j}}{K_{i+1}-K_{i}}
\]

We may interpret each $q_{i,j}$ as the marginal risk-neutral probability
that the stock price at maturity $T_{j}$ equals $K_{i}$. 

For future use, we associate with each maturity a risk-neutral probability
measure defined by
\[
\mathbb{Q}_{j}(K)=\sum_{K_{j}\leq K}q_{i,j}
\]

A third test on the provided call quotes requires that for each discrete
strike $K_{i},i\geq0$, and each discrete maturity $T_{j},j\geq0,$we
have 
\begin{equation}
C_{i,j+1}-C_{i,j}\geq0\label{eq:thirdTest}
\end{equation}

The left-hand side of \eqref{eq:thirdTest} is the cost of a calendar
spread consisting of long one call of maturity $T_{j+1}$ and short
one call of maturity $T_{j}$, with both calls struck at $K_{i}$.
Hence, our third test requires that calendar spreads comprised of
adjacent maturity calls are not negatively priced at each maturity.

We now conclude, following \cite{Carr_Madan_2005} , the discussion
on the 3 arbitrage constraints \eqref{eq:firstTest}\eqref{eq:secondTest}\eqref{eq:thirdTest}.

As the call pricing functions are linear interpolations of the provided
quotes, we have that at each maturity $T_{j}$, calendar spreads are
not negatively priced for the continuum of strikes $K>0$. Since all
convex payoffs may be represented as portfolios of calls with non-negative
weights, it follows that all convex functions $\phi(S)$ are priced
higher when promised at $T_{j+1}$ than when they are promised at
$T_{j}$. In turn, this ordering implies that the risk-neutral probability
measures $\mathbb{Q}$ constructed above are increasing in the convex
order with respect to the index $j$. This implies that there exists
a martingale measure which is consistent with the call quotes and
which is defined on some filtration that includes at least the stock
price and time. Finally, it follows that the provided call quotes
are free of static arbitrage by standard results in arbitrage pricing
theory.

\section{Characteristics of volatility surface}

Many recent papers have studied various characteristics of volatility
surface:
\begin{itemize}
\item the static and dynamic properties of the implied volatility surface
must exhibit within an arbitrage-free framework
\item implied volatility calculations in a (local) stochastic volatility
environment, which may also include jumps or even Levy processes. 
\item the behavior of implied volatility in limiting cases, such as extreme
strikes, short and large maturities, etc.
\end{itemize}
For completion we include a list of relevant papers: \cite{Ayache_et_al_2004,Benaim_Friz_2009,Berestycki_et_al_2004,Cheng_et_al_2009,Cont_daFonseca_2002,Cousot_2007,Daglish_et_al_2007,Davis_Hobson_2007,Decamps_DeSchepper_2008,Durrleman_2010,FigueroaLopez_Forde_2011,Forde_et_al_2010,Forde_et_al_2011,Forde_Jacquier_2009,Forde_Jacquier_2009a,Forde_Jacquier_2009b,Forde_Jacquier_2011,Forde_Jacquier_2011a,Friz_et_al_2011,Fukasawa_2011,Gao_Lee_2011,Gatheral_2006,Gatheral_et_al_2011,Gulisashvili_2010,Henry-Labordere_2005,HenryLabordere_2008,Jacquier_2010,Laurence_2008,Lee_2004,Lee_2004a,Lipton_2002,Medvedev_2008,Medvedev_Scaillet_2006,Obloj_2008,Osajima_2007,Paulot_2009,Reghai_2010,Reiswich_2010,Roper_2010,Roper_2010a,Takahashi_2007,Takahashi_et_al_2009,Takahashi_et_al_2011,Takehara_et_al_2010,Taylor_2011,Tehranci_2009,Benhamou_et_al_2008,Benhamou_et_al_2009,Benhamou_et_al_2011,Bergomi_2004,Bergomi_2005,Bergomi_2008,Bergomi_2009,Ayache_2005,Durrleman_ElKaroui_2007,Friz_2005,Galluccio_LeCam_2007,Jacquier_2010a,KellerRessel_2011,Takahashi_Yamada_2009,Wang_2007,Benaim_Friz_2009a}

The constructed volatility surface may also need to take into account
the expected behavior of the volatility surface outside the core region.
The core region is defined as the region of strikes for equity markets,
moneyness levels for Commodity markets, deltas for FX markets for
which we have observable market data. From a theoretical point of
view, this behavior may be described by the asymptotics of implied
volatility, while from a practical point of view this corresponds
to smile extrapolation.

\subsection{Asymptotics of implied volatility}

Concerning the dependence with respect to strike, some major theoretical
results are known in a model-independent framework. \cite{Lee_2004}
related the extreme strike slopes of the implied volatility to the
critical moments of the underlying through the moment formula: let
$\sigma(t,x)$ denote the implied volatility of a European Call option
with maturity $t$ and strike $K=F_{0}e^{x}$, then
\begin{equation}
\lim_{x\rightarrow}\sup_{\infty}\frac{t\sigma\left(t,x\right)^{2}}{x}=\psi\left(u^{*}(t)-1\right)\label{eq:lee}
\end{equation}

where $\psi(u)=2-4\left(\sqrt{u^{2}+u}-u\right)$ and $u^{*}\left(t\right)\triangleq\sup\left\{ u\geq1;E\left[F^{u}(t)\right]\right\} $
is the critical moment of the underlying price $F=\left(F(t)\right)_{t\geq0}$.
An analogous formula holds for the left part of the smile, namely
when $x\rightarrow-\infty$. This result was sharpened in \cite{Benaim_Friz_2009,Benaim_Friz_2009a},
relating the left hand side of \eqref{eq:lee} to the tail asymptotics
of the distribution of $F(t)$.

In the stochastic volatility framework this formula was applied by
\cite{Andersen_Piterbarg_2007} and \cite{KellerRessel_2011}, to
mention but a few.

The study of short- (resp. long-) time asymptotics of the implied
volatility is motivated by the research of efficient calibration strategies
to the market smile at short (resp. long) maturities. Short time results
have been obtained in \cite{Medvedev_Scaillet_2006,Forde_Jacquier_2009,Forde_Jacquier_2009a,Forde_Jacquier_2009b,Berestycki_et_al_2004},
while some other works provide insights on the large-time behavior,
as done by \cite{Tehranci_2009} in a general setting, \cite{KellerRessel_2011}
for affine stochastic volatility models or \cite{Forde_Jacquier_2011}
for Heston model.

\subsection{Smile extrapolation}

It is argued in the practitioner paper \cite{Benaim_et_al_2009} that
a successful smile extrapolation method should deliver arbitrage-free
prices for the vanilla options, i.e., the option prices must be convex
functions of strike, and stay within certain bounds. In addition,
the extrapolation method should ideally have the following properties: 
\begin{enumerate}
\item It should reprice all observed vanilla options correctly. 
\item The PDF, CDF and vanilla option prices should be easy to compute. 
\item The method should not generate unrealistically fat tails, and if possible,
it should allow us to control how fat the tails are. 
\item It should be robust and flexible enough to use with a wide variety
of different implied volatility surfaces. 
\item It should be easy and fast to initialize for a given smile. 
\end{enumerate}
The paper describes two commonly used methods which do not satisfy
the above wish list. The first one is to use some interpolation within
the region of observed prices, and just set the implied volatility
to be a constant outside of this region. This method is flawed as
it introduces unstable behavior at the boundary between the smile
and the flat volatility, yielding unrealistically narrow tails at
extreme strikes. 

The second approach is to use the same parametric form for the implied
volatility derived from a model, e.g. SABR, both inside and outside
the core region. There are several problems with this method. It gives
us little control over the distribution; indeed this approach often
leads to excessively fat tails, which can lead to risk neutral distributions
that have unrealistically large probabilities of extreme movements,
and have moment explosions that lead to infinite prices, even for
simple products. If the methodology is dependent on usage of an asymptotic
expansion, the expansion may become less accurate at strikes away
from the money, leading to concave option prices, or equivalently
negative PDFs, even at modestly low strikes. Furthermore, there is
no guarantee that this functional form will lead to arbitrage free
prices for very large and small strikes. 

That is why \cite{Benaim_et_al_2009} propose to separate the interpolation
and extrapolation methods. 

Their method works as follows. A core region of observability, inside
which we may use any standard smile interpolation method, is defined
first: $K_{-}\leq K\leq K_{+}$. Outside of this region the extrapolation
is done by employing a simple analytical formula for the option prices,
that has the following characteristics:
\begin{itemize}
\item For very low strikes region, the formula-based put prices will go
towards zero as the strike goes to zero, while remaining convex.
\item For very high strikes region, the formula-based call prices will go
towards zero as strike goes to infinity, while remaining convex.
\end{itemize}
Each of these formulas is parametrized so that we can match the option
price as well as its first two derivatives at the corresponding boundary
with the core region. The methodology is also able to retain a measure
of control over the form of the tails at extreme strikes.

The following functional form for the extrapolation of put and, respectively,
call prices was described as parsimonious yet effective:
\begin{eqnarray*}
Put(K) & =K^{\mu} & \exp\left[a_{1}+b_{1}K+c_{1}K^{2}\right]\\
Call(K) & =K^{-\nu} & \exp\left[a_{2}+\frac{b_{2}}{K}+\frac{c_{2}}{K^{2}}\right]
\end{eqnarray*}

We fix $\mu$>1, which ensures that the price is zero at zero strike,
and there is no probability for the underlying to be zero at maturity.
Alternatively, we can choose $\mu$ to reflect our view of the fatness
of the tail of the risk neutral distribution. It is easy to check
that this extrapolation generates a distribution where the $m$-th
negative moment is finite for $m<1-\mu$ and infinite for $m>1-\mu$. 

We fix $\nu>0$ to ensure that the call price approaches zero at large
enough strikes. Our choice of controls the fatness of the tail; the
$m$-th moment will be finite if $m<\nu-1$ and infinite if $m>\nu-1$.

The condition for matching the price and its first two derivatives
at $K_{-}$ and, respectively, at $K_{+}$ yields a set of linear
equations for the parameters $a_{1},b_{1},c_{1}$ and, respectively,
for $a_{2},b_{2},c_{2}$

\section{Remarks on numerical calibration}

The calibration procedure consists of finding the set of parameters
(defining the volatility surface) that minimize the error between
model output and market data, while satisfying some additional constraints
such as {}``no arbitrage in strike and time'', smoothness, etc.
This chapter provides some details regarding the practical aspects
of numerical calibration.

Let us start by making some notations: we consider that we have $M$expiries
$\left\{ T^{(j)}\right\} $ and that for each maturity $T^{(j)}$
we have $N[j]$ calibrating instruments, with strikes $K_{i,j}$,
for which market data is given (as prices or implied volatilities).
The bid and ask values are denoted by $Bid\left(i,T^{(j)}\right)$
and $Ask\left(i,T^{(j)}\right)$

\subsection{Calibration function }

The calibration function is defined in different ways

If we perform {}``all-at-once'' calibration, then the calibration
function is constructed as

\[
\Psi\triangleq\sum_{j=1}^{M}\sum_{i=1}^{N[j]}w_{i,j}\left\Vert ModelOutput\left(i,T^{(j)}\right)-MarketData\left(i,T^{(j)}\right)\right\Vert 
\]

where $w_{i,j}$ are weights and $\left\Vert \cdot\right\Vert $ denotes
a generic norm

If we perform sequential calibration, one expiry at the time, than
the calibration functional for each expiry will be given as

\[
\Psi\left[j\right]\triangleq\sum_{i=1}^{N[j]}w_{i,j}\left\Vert ModelOutput\left(i,T^{(j)}\right)-MarketData\left(i,T^{(j)}\right)\right\Vert 
\]

If we use a local optimizer, then we might need to add a regularization
term. The regularization term the most commonly considered in the
literature is of Tikhonov type. e.g., \cite{Orosi_2010a,Cont_Tankov_2006,Maruhn_2011,Achdou_Pironneau_2005}.
However, since this feature is primarily employed to ensure that the
minimizer does not get stuck in a local minimum, this additional term
is usually not needed if we use either a global optimizer or a hybrid
(combination of global and local) optimizer.

\subsection{Constructing the weights in the calibration functional}

The weights $w_{i,j}$ can be selected following various procedures
detailed in \cite{Bloch_2010,Bloch_Coello_Coello_2010,Cont_Tankov_2002,Crosby_2010,Lindstrom_et_al_2008}
chapter 13 of \cite{Cont_Tankov_2004}, to mention but a few.

Practitioners usually compute the weights (see \cite{Crosby_2010})
as inverse proportional to
\begin{itemize}
\item the square of the bid-ask spreads, to give more importance to the
most liquid options.
\item the square of the $\bsm$ Vegas (roughly equivalent to using implied
volatilities, as explained below). 
\end{itemize}
\cite{Lindstrom_et_al_2008} asserts that it is statistically optimal
(minimal variance of the estimates) to choose the weights as the inverse
of the variance of the residuals, which is then considered to be proportional
to the inverse of squared bid\textendash{}ask spread. 

Another practitioner paper \cite{Bloch_2010} considers a combination
of the 2 approaches and this is our preferred methodology. 

It is known that at least for the options that are not too far from
the money, the bid-ask spreads is of order of tens of basis points.
This suggests that it might be better to minimize the differences
of implied volatilities instead of those of the option prices, in
order to have errors proportional to bid-ask spreads and to have better
scaling of the cost functional. However, this method involves additional
computational cost. A reasonable approximation is to minimize the
square differences of option prices weighted by the $\bsm$ Vegas
evaluated at the implied volatilities of the market option prices.

The starting point is given by setting the weights as
\[
w_{i,j}=\frac{1}{\left|Bid\left(i,T^{(j)}\right)-Ask\left(i,T^{(j)}\right)\right|}
\]

To simplify the notations for the remainder of the chapter we denote
the model price by $ModP$ and the market price by $MktP$

We can approximate the difference in prices as follows: 
\[
ModP\left(i,T^{(j)}\right)-MktP\left(i,T^{(j)}\right)\thickapprox\part{\left[MktP\left(i,T^{(j)}\right)\right]}{\sigma_{IV}}\left(\sigma_{IV}^{MOD}\left(i,T^{(j)}\right)-\sigma_{IV}^{MKT}\left(i,T^{(j)}\right)\right)
\]

where $\sigma_{IV}^{MOD}\left(i,T^{(j)}\right)$ and $\sigma_{IV}^{MKT}\left(i,T^{(j)}\right)$
are the implied vols corresponding to model and, respectively, market
prices for strikes $K_{i,j}$ and maturities $T^{(j)}$.

We should note that this series approximation may be continued to
a higher order if necessary, with a very small additional computational
cost.

Using the following expression for $\bsm$ Vegas evaluated at the
implied volatility $\sigma_{IV}^{MKT}\left(i,T^{(j)}\right)$ of the
market option prices
\[
\part{\left[MktP\left(i,T^{(j)}\right)\right]}{\sigma_{IV}}=Vega\left(\sigma_{IV}^{MKT}\left(i,T^{(j)}\right)\right)
\]

we obtain 
\[
\sigma_{IV}^{MOD}\left(i,T^{(j)}\right)-\sigma_{IV}^{MKT}\left(i,T^{(j)}\right)\thickapprox\frac{1}{Vega\left(\sigma_{IV}^{MKT}\left(i,T^{(j)}\right)\right)}\left[ModP\left(i,T^{(j)}\right)-MktP\left(i,T^{(j)}\right)\right]
\]

Thus we can switch from difference of implied volatilities to difference
of prices. For example, for all-at-once calibration that is based
on minimization of root mean squared error (RMSE) , the corresponding
calibration functional can be written as 
\[
\Psi\triangleq\sum_{j=1}^{M}\sum_{i=1}^{N[j]}\bar{w}_{i,j}\left[ModelOutput\left(i,T^{(j)}\right)-MarketData\left(i,T^{(j)}\right)\right]^{2}
\]

where the weights $\bar{w}_{i,j}$ are defined as
\[
\bar{w}_{i,j}=\frac{1}{\left|Bid\left(i,T^{(j)}\right)-Ask\left(i,T^{(j)}\right)\right|}\left(\frac{1}{Vega\left(\sigma_{IV}^{MKT}\left(i,T^{(j)}\right)\right)}\right)^{2}
\]

To avoid overweighting of options very far from the money we need
introduce an upper limit for the weights.

\subsection{Selection of numerical optimization procedure}

It is quite likely that the calibration function for the volatility
surface may exhibit several local (and perhaps global) minima, making
standard optimization techniques somewhat unqualified according to
\cite{Cont_BenHamida_2005}. Gradient based optimizers, for example,
are likely to get stuck in a local minimum which may also be strongly
dependent on the initial parameter guess. While this situation (multiple
local minima) may be less common for equity markets , it is our experience
that such characteristics are quite common for FX and Commodities
markets. Thus global/hybrid optimization algorithms are our preferred
optimization methods in conjunction with volatility surface construction. 

Our favorite global optimization algorithm is based on Differential
Evolution \cite{Price_et_al_2005}. In various flavors it was shown
to outperform all other global optimization algorithms when solving
benchmark problems (unconstrained, bounded or constrained optimization).
Various papers and presentations described successful calibrations
done with Differential Evolution in finance: \cite{Bloch_2010,Bloch_Coello_Coello_2010,Detlefsen_Hardle_2008,Vollrath_Wendland_2009,Cont_BenHamida_2005},
to mention but a few.

Here is a short description of the procedure. Consider a search space
$\Theta$ and a continuous function $G:\Theta\rightarrow\mathbb{R}$
to be minimized on $\Theta$. An evolutionary algorithm with objective
function G is based on the evolution of a population of candidate
solutions, denoted by $X_{n}^{N}=\left\{ \theta_{n}^{i}\right\} _{i=1...N}$.
The basic idea is to {}``evolve'' the population through cycles
of modification (mutation) and selection in order to improve the performance
of its individuals. At each iteration the population undergoes three
transformations: 
\[
X_{n}^{N}\rightarrow V_{n}^{N}\rightarrow W_{n}^{N}\rightarrow V_{n+1}^{N}
\]

During the mutation stage, individuals undergo independent random
transformations, as if performing independent random walks in $\Theta$,
resulting in a randomly modified population $V_{n}^{N}$. In the crossover
stage, pairs of individuals are chosen from the population to \textquotedbl{}reproduce\textquotedbl{}:
each pair gives birth to a new individual, which is then added to
the population. This new population, denoted $W_{n}^{N}$, is now
evaluated using the objective function $G\left(\cdot\right)$. Elements
of the population are now selected for survival according to their
fitness: those with a lower value of $G$ have a higher probability
of being selected. The $N$ individuals thus selected then form the
new population $X_{n+1}^{N}$. The role of mutation is to explore
the parameter space and the optimization is done through selection.

On the downside, global optimization techniques are generally more
time consuming than gradient based local optimizers. Therefore, we
employ a hybrid optimization procedure of 2 stages which combines
the strengths of both approaches. First we run a global optimizer
such as Differential Evolution for a small number of iterations, to
arrive in a {}``neighborhood'' of a global minimum. In the second
stage we run a gradient-based local optimizer, using as initial guess
the output from the global optimizer, which should converge much quite
fast since the initial guess is assumed to be in the {}``correct
neighborhood''. An excellent resource for selecting a suitable local
optimizer can be found at \cite{OptimizSoftware}

We should also mention that a very impressive computational speedup
(as well as reducing number of necessary optimization iterations)
can be achieved if the gradient of the cost functional is computed
using Adjoint method in conjunction with Automatic, or Algorithmic,
Differentiation (usually termed AD). Let us exemplify the computational
savings. Let us assume that the calibration functional depends on
P parameters, and that the computational cost for computing one instance
of the calibration functional is T time units. The combination between
adjoint and AD methodology is theoretically guaranteed to produce
the gradient of of the calibration functional (namely all P sensitivities
with respect to parameters) in a computational time that is not larger
than 4-5 times the original time T for computing one instance of the
calibration functional. The gradient is also very accurate, up to
machine precision, and thus we eliminate any approximation error that
may come from using finite difference to compute the gradient. The
local optimizer can then run much more efficiently if the gradient
of the calibration functional is provided explicitly. For additional
details on adjoint plus AD the reader is referred to \cite{Homescu_2011}

\section{Conclusion}

We have surveyed various methodologies for constructing the implied
volatility surface. We have also discussed related topics which may
contribute to the successful construction of volatility surface in
practice: conditions for arbitrage/non arbitrage in both strike and
time, how to perform extrapolation outside the core region, choice
of calibrating functional and selection of optimization algorithms.

\bibliographystyle{plain}
\phantomsection\addcontentsline{toc}{section}{\refname}\bibliography{References_VolSurface_Arxiv}

\end{document}